\documentclass[12pt,a4paper]{iopart}
\usepackage{iopams}
\usepackage{graphicx}
\usepackage{setstack,undertilde}
\usepackage[colorlinks=true,linkcolor=blue,citecolor=blue,filecolor=blue,urlcolor=blue]{hyperref}
\def\v{\boldsymbol v}
\def\s{\boldsymbol s}
\def\bb{\boldsymbol b}

\def\br{\boldsymbol r}
\def\bu{\boldsymbol u}

\def\fh{\hat{f}}

\def\Real{\mathrm{Re}}

\def\OO{\mathrm{O}}
\def\U{\mathrm{U}}

\def\tW{\tau_{\mathrm{W}}}
\def\tH{T_{\mathrm{H}}}

\def\btW{\bar{\tau}_{\mathrm{W}}}
\def\var{\mathrm{var}}
\def\Heff{\mathcal{H}_\mathrm{eff}}

\newcommand{\ie}{\textit{i.e.}\ }
\newcommand{\eg}{\textit{e.g.}\ }
\newcommand{\cf}{\textit{c.f.}\ }

\newcommand{\frefs}[1]{figures~\ref{#1}}

\begin{document}

\title[Efficient semiclassical approach for time delays]{Efficient semiclassical approach for time delays}

\author{Jack Kuipers$^1$, Dmitry V.\ Savin$^2$ and Martin Sieber$^3$}
\address{$^1$Institut f\"ur Theoretische Physik, Universit\"at Regensburg, D-93040
Regensburg, Germany}
\address{$^2$Department of Mathematics, Brunel University London, Uxbridge, UB8 3PH, United Kingdom}
\address{$^3$School of Mathematics, University of Bristol, Bristol, BS8 1TW, United Kingdom}
\ead{jack.kuipers@ur.de}

\begin{abstract}
The Wigner time delay, defined by the energy derivative of the total scattering phase shift, is an important spectral measure of an open quantum system characterising the duration of the scattering event. It is proportional to the trace of the Wigner-Smith matrix $Q$ that also encodes other time-delay characteristics. For chaotic cavities, these quantities exhibit universal fluctuations that are commonly described within random matrix theory. Here, we develop a new semiclassical approach to the time-delay matrix which is formulated in terms of the classical trajectories that connect the exterior and interior regions of the system.  This approach is superior to previous treatments because it avoids the energy derivative. We demonstrate the method's efficiency by going beyond previous work in establishing the universality of time-delay statistics for chaotic cavities with perfectly connected leads. In particular, the moment generating function of the proper time-delays (eigenvalues of $Q$) is found semiclassically for the first five orders in the inverse number of scattering channels for systems with and without time-reversal symmetry. We also show the equivalence of random matrix and semiclassical results for the second moments and for the variance of the Wigner time delay at any channel number.
\end{abstract}

\setlength{\unitlength}{1cm}
\begin{picture}(10,0)(-3,-11.5)
 \put(-3.75,-15){\parbox{0.75\textwidth}{\small Published 8 December 2014
 \\[1ex] \emph{New Journal of Physics} \textbf{16} (2014) 123018}}
\end{picture}
\vspace*{-2ex}

\pacs{05.45.Mt, 03.65.Sq, 03.65.Nk, 73.23.-b}
%% 05.45.Mt (Quantum chaos; semiclassical methods)
%% 03.65.Sq  (Semiclassical theories and applications)
%% 03.65.Nk (Scattering theory)
%% 73.23.-b (Electronic transport in mesoscopic systems)
\vspace{2pc}
{\small\noindent{\it Keywords}: semiclassical approach, random matrix theory, Wigner time delay}
%\submitto{\NJP}
\maketitle

%\markboth{}{}
%\tableofcontents
%\markboth{}{}

\section{Introduction}\label{intro}

The concept of time delay plays an important and special role in quantum collision theory. It was first introduced by Eisenbud and Wigner \cite{eisenbud48,wigner55} in the context of elastic scattering, who related the time during which a monochromatic wave packet is delayed in the interaction region to the energy derivative of the scattering phase shift. Later on Smith \cite{smith60} extended the concept to the case of inelastic scattering with many channels and introduced the time-delay (or Wigner-Smith) matrix
\begin{equation} \label{eq:Q}
\fl Q(E) = -\rmi\hbar S^{\dagger}(E) \frac{\rmd S(E)}{\rmd E}
  = -\rmi\hbar\frac{\rmd}{\rmd \epsilon} \left[ S^{\dagger}\left(E-\frac{\epsilon}{2}\right) S\left(E+\frac{\epsilon}{2}\right) \right] \Big\vert_{\epsilon=0} \, ,
\end{equation}
where $S(E)$ is the scattering matrix at energy $E$ whose dimension is given by the number of open channels, $M$. For an energy and flux conserving scatterer (\ie involving no absorption or dissipation), the $S$-matrix is unitary and $Q$ is a Hermitian matrix, which is manifest from the second representation in \eref{eq:Q}. The real characteristics of $Q$ (\eg  its diagonal elements and eigenvalues) provide us with various time-delay related observables \cite{smith60}. Taking the trace, one arrives at the simple weighted-mean measure of the collision duration  \cite{lyuboshitz77,lewe91,lehmannetal95,fs97}
\begin{equation} \label{eq:tW}
 \tW(E) \equiv \frac{1}{M}\Tr Q(E) = - \frac{\rmi\hbar}{M} \frac{\rmd }{\rmd E}\ln\det S(E) \, .
\end{equation}
This expression defines the Wigner time delay in multi-channel scattering. Further and more subtle discussions of the physical concepts of time delays and their applications to regular and chaotic scattering can be found in recent reviews \cite{fs97,muga00,deCar02}.

As is clear from the definition, the dominant contributions to $\tW$ come from the regions where the total scattering phase shift exhibits a sharp energy dependence. Away from the channel thresholds, this occurs in the vicinity of resonances that correspond to the meta-stable (decaying) states formed at the intermediate stage of the scattering event. Such resonances can be analytically viewed as the complex poles $\mathcal{E}_n=E_n-\frac{\rmi}{2}\Gamma_n$ of the $S$-matrix in the complex energy plane, with $E_n$ and $\Gamma_n$ being the position and width of the $n$th resonance, respectively. In the resonance approximation, neglecting the constant background connected to potential scattering and direct reactions, these poles are the only singularities of $S$. In view of the unitarity condition, their complex conjugates $\mathcal{E}^*_n$ serve as the $S$-matrix zeros, thus implying
$\det S =\prod_n\frac{E-\mathcal{E}^*_n}{E-\mathcal{E}_n}$. This results in the important connection \cite{lyuboshitz77,lehmannetal95}
\begin{equation} \label{eq:tWdens}
 \tW(E) = \frac{\hbar}{M}\sum_n\frac{\Gamma_n}{(E-E_n)^2+\frac{1}{4}\Gamma_n^2} \, ,
\end{equation}
between the Wigner time delay and the resonance spectrum of the intermediate open system. This expression is valid at arbitrary degree of the resonance overlap and thus the Winger time delay \eref{eq:tW} can also be treated as the density of states of the open system, see \cite{friedel52,krei53} for relevant discussion. In particular, the spectral average of $\tW$  over a narrow energy window is $\bar\tW=2\pi\hbar\bar\rho/M$, where $\bar\rho$ is the mean density around $E$. The latter in turn determines the fundamental timescale in quantum systems, the Heisenberg time $\tH=2\pi\hbar\bar\rho$, so that $\bar\tW=\tH/M$.

In many experimental situations, the intermediate system is characterised by very complicated internal motion, with representative examples being microwave cavities \cite{Stoeckmann}, quantum dots \cite{alha00} and compound nuclei \cite{mitc10}. As a result, the time delay (as well as other scattering observables) reveals strong fluctuations around its mean value that need to be treated statistically. There are two main approaches to describe such fluctuations: random matrix theory (RMT) \cite{Mehta2} and the semiclassical method \cite{gutzwiller90}.

The general RMT approach to the time delay, mostly developed in \cite{lehmannetal95,fs97,sss01}, is based on a representation of the Wigner-Smith matrix in terms of the effective non-Hermitian Hamiltonian $\Heff$ of the open system. The complex resonances $\mathcal{E}_n$ are then given by the eigenvalues of the Hamiltonian \cite{soko89}. The Hermitian part of $\Heff$ describes the internal Hamiltonian with a discrete spectrum and chaotic dynamics. It can therefore be modelled by a random $N{\times}N$ matrix drawn from the Gaussian orthogonal (GOE) or unitary (GUE) ensemble, depending on the presence or absence of time-reversal symmetry (TRS), respectively \cite{Mehta2}. The anti-Hermitian part originates from coupling between the $N$ internal and $M$ channels states and has a specific algebraic structure due to the unitarity constraint \cite{soko89}. The statistical averaging can then be performed by making use of the powerful supersymmetry technique, mapping the problem to a nonlinear supersymmetric $\sigma$-model of zero-dimensional field theory \cite{verb85}. In the asymptotic limit $N\to\infty$, the obtained results are universal in the sense that local fluctuations on the scale of mean level spacing $1/\bar{\rho}\sim N^{-1}$ become model-independent provided that the appropriate natural units have been used \cite{guhr98}. The only parameters are then the number $M$ of open channels and their strength of coupling to the continuum. The later is quantified by the average `optical' $S$-matrix, with $\overline{S}=0$ (or $\overline{S}\to1$) being the limiting case of a perfectly open (or almost closed) system.

The exact universal results were initially obtained for the autocorrelation function of the Wigner time delay, first for orthogonal \cite{lehmannetal95} and unitary symmetry \cite{fs97}, and then for the whole crossover of partly broken TRS \cite{fyod97a}. Along these lines, exact formulae were also derived for the distribution function of the \emph{partial} time-delays \cite{fs97,fyod97a} (defined by the energy derivatives of the $S$-matrix eigenphases and related to the diagonal elements of $Q$ \cite{savi01}) as well as that of the \emph{proper} time-delays (given by the eigenvalues of $Q$) \cite{sss01,ss03}. The developed method is actually flexible enough to also include the effects of finite absorption \cite{ss03,fyod05} and disorder \cite{fyod05,ossi05}. The most recent overview of the relevant results can be found in \cite{fyod11ox}.

Further progress is possible in the particular case of perfect coupling, when $S$ is uniformly distributed in Dyson's circular ensemble of the appropriate symmetry \cite{beenakker97}. Brouwer et al.\ \cite{bfb97} exploited the invariance properties of the energy-dependent $S$-matrix in this case to derive the distribution of the whole time-delay matrix, generalising the earlier one-channel result \cite{gopa96} to arbitrary $M$. The joint distribution function of its eigenvalues, \ie the proper time-delays, turned out to be determined by the Laguerre ensemble from RMT \cite{bfb97}. This provided a route to apply orthogonal polynomials to compute marginal distributions \cite{bfb99} and various moments \cite{mezz11,mezz12,mart14,nova14} or use a Coulomb gas method to study the total density \cite{texi13}. Very recently, Mezzadri and Simm \cite{mezz13} applied the integrable theory of certain matrix integrals to the problem and developed an efficient method for computing cumulants of arbitrary order. In particular, the variance of the Wigner time delay was found in the simple form,
\begin{equation}\label{eq:tWvar}
 \var(\tW) \equiv \frac{1}{\btW^2}\left\langle \left(\tW-\btW\right)^2 \right\rangle=\frac{4}{(M+1)(\beta M-2)} \, ,
\end{equation}
where $\langle \cdots \rangle$ stands for a statistical average and $\beta=1$ ($\beta=2$) indicates orthogonal (unitary) symmetry. This expression agrees with earlier results obtained at arbitrary coupling in \cite{lehmannetal95,fs97}. We note, however, that the results for the distribution of Wigner time delay are still rather limited, with explicit expressions being available only at $M=1,2$ \cite{gopa98,savi01} or in the limit of $M\gg1$ \cite{texi13}.

Taking a semiclassical approach, an early expression for the Wigner time delay was derived through the connection to the density of states mentioned above. The fluctuating part of this quantity can be expressed, like Gutzwiller's trace formula for closed systems \cite{gutzwiller71,gutzwiller90}, as a sum over periodic orbits,
\begin{equation} \label{eq:tWsemipos}
\tW-\btW\approx\frac{2}{M}\Real \sum_{p,m} A_{p,m} \rme^{\frac{\rmi}{\hbar}m\mathcal{S}_p} \, ,
\end{equation}
but now only involving those orbits trapped inside the scattering system \cite{bb74}. The trapped primitive orbits $p$ and their repetitions $m$ involve their actions $\mathcal{S}_{p}$ and stability amplitudes $A_{p,m}$ which in turn depend on the period, monodromy matrix and Maslov index of the orbits. Since the dependence on the energy (and other parameters) is then encoded in the sum over the periodic orbits, this can be used to calculate correlation functions \cite{eckh93,val98}, in particular, the time-delay variance is expressed as
\begin{equation} \label{eq:timedelayvar}
 \var(\tW) \approx \frac{2}{\tH^2} \Real \left\langle
  \sum_{p,p'} A_{p}A_{p'}^* \rme^{\frac{\rmi}{\hbar}(\mathcal{S}_p-\mathcal{S}_{p'})} \right\rangle \, .
\end{equation}
When taking the semiclassical limit $\hbar\to0$, the rapid oscillations in the phase induced will be washed out by the overall spectral averaging unless there
are systematic correlations between periodic orbits with action difference $\vert\mathcal{S}_p-\mathcal{S}_{p'}\vert \lesssim \hbar$.
The repetitions have been ignored in \eref{eq:timedelayvar} since they are exponentially smaller than the $m=1$ term. The semiclassical treatment of sums like in \eref{eq:timedelayvar} then involves identifying and evaluating correlated pairs of orbits with a small action difference. The simplest pair is to set $p=p'$, known as the diagonal approximation \cite{berry85}, which can be evaluated using the open system version \cite{ce91} of the Hannay-Ozorio de Almeida sum rule \cite{ha84}
\begin{equation}
 \sum_{p}\vert A_{p} \vert^2 \approx \int_{0}^{\infty} t \rme^{-\mu t} \rmd t = \frac{1}{\mu^2} \, .
\end{equation}
Here the exponential weighting corresponds to the expected number of periodic orbits of the closed system which survive up to a time $t$, with $\mu=M/\tH$ being the classical escape rate (see, however, the discussion in \cite{lehmannetal95,ss97,lv04}). With TRS, one can also pair an orbit with its time reverse and obtain a further factor of 2, or
\begin{equation} \label{eq:vartdleadingord}
 \var(\tW) = \frac{4}{\beta M^2} + O(M^{-3}) \, ,
\end{equation}
thus reproducing \eref{eq:tWvar} to leading order. This is not surprising, of course, as the quasiclassical limit in the RMT treatment corresponds to the asymptotic case of $M\gg1$ \cite{lewe91,lehm95a}. The quantum effects are encoded in the higher order terms of the $1/M$ expansion, being in general responsible for slowing down the decay law in chaotic systems from the purely exponential one \cite{ss97}. It is now well understood that the higher order terms can be obtained semiclassically through a systematic expansion of correlated periodic pairs which was first derived for the spectral statistics of closed systems \cite{sr01,mulleretal04,mulleretal05}.
The extension to open systems and the time delay was then developed in \cite{ks07b}, providing the terms up to $M^{-8}$ in full agreement with the RMT result. Further calculations along those lines become too involved because of the quickly growing complexity of relevant combinatorics. One possibility to overcome such difficulties might be in doing the semiclassical approximation at the level of the generating functions \cite{heus07} used to derive the corresponding $\sigma$-model of RMT \cite{lehmannetal95,fs97}, the idea that has already proven to be success in establishing the full equivalence with RMT for the closed systems \cite{muel09}.
However, we will exploit another route here.

An alternative starting point relies on the van Vleck approximation of the propagator that gives the semiclassical approximation for the scattering matrix elements \cite{miller75,jbs90,bjs93b,richter00} in terms of trajectories that connect the corresponding (say, input $i$ and output $o$) channels
\begin{equation}
S_{oi} \approx \frac{1}{\sqrt{\tH}}\sum_{\gamma(i\to o)} \tilde{A}_{\gamma} \rme^{\frac{\rmi}{\hbar}\mathcal{S}_\gamma} \, .
\end{equation}
The scattering trajectories now have different stability amplitudes $\tilde{A}_{\gamma}$ which do not explicitly depend on the duration $T_{\gamma}$ of the trajectories but still include a phase factor.
Taking the energy derivative and noting that $\partial\mathcal{S}_{\gamma} /\partial E = T_{\gamma}$,
one obtains the approximation for the Wigner time delay \cite{lv04}
\begin{equation}\label{eq:tWsemi}
 \tW \approx \frac{1}{M\tH} \sum_{i,o =1}^{M} \sum_{\gamma,\gamma'(i\to o)} T_{\gamma} \tilde{A}_{\gamma} \tilde{A}_{\gamma'}^* \rme^{\frac{\rmi}{\hbar}(\mathcal{S}_{\gamma}-\mathcal{S}_{\gamma'})} \, ,
\end{equation}
where we have assumed that the amplitudes vary much more slowly than the phase. The diagonal approximation $\gamma=\gamma'$ can again be evaluated with a sum rule \cite{rs02}
\begin{equation} \label{eq:diagscattermeantd}
 \btW \approx \frac{1}{M\tH} \left\langle \sum_{i,o =1}^{M} \sum_{\gamma(i\to o)} T_{\gamma}\vert\tilde{A}_{\gamma} \vert^{2}\right\rangle \approx \frac{M}{\tH} \int_{0}^{\infty} t \rme^{-\mu t} \rmd t = \frac{1}{\mu} \ ,
\end{equation}
which directly gives the average time delay.

In the presence of TRS one can also partner a scattering trajectory with its time reverse if they start and end in the same channel. However there are further correlated pairs of trajectories, both with and without TRS. They can be related to correlated pairs of periodic orbits by formally cutting the periodic orbits open and  deforming them into scattering trajectories \cite{rs02,heusleretal06,mulleretal07}. It is worth stressing that the second form of the time-delay matrix in \eref{eq:Q} turns out to be particularly useful for the semiclassical treatment. One can show in this way \cite{ks08} that all further contributions to the average time delay cancel leaving only the diagonal terms intact. Furthermore, one can also derive \eref{eq:tWsemipos} semiclassically from \eref{eq:tWsemi} by recreating the sum over periodic orbits in \eref{eq:tWsemipos} from correlations of scattering trajectories that approach the trapped periodic orbits and follow them for many repetitions. This provides a duality between the two approaches and allows access to a range of statistics beyond the average time delay \cite{ks08}. For example, the moments of the proper time-delays can be found to leading order in $M^{-1}$ by using energy-dependent correlators of the $S$-matrix elements and performing a semiclassical expansion for the later \cite{bk10}. The next two orders can similarly be obtained using a recursive graphical representation of the semiclassical diagrams of correlated trajectories \cite{bk11}, showing an agreement with RMT \cite{mezz12}, though the energy dependence, which is later differentiated out and removed as in \eref{eq:Q}, complicates the treatment drastically.

In this paper, we derive yet another semiclassical approximation to the time delay which avoids using such an energy differentiation in the first place and significantly simplifies the semiclassical calculations.
It builds upon the resonant representation \cite{sz97} of the matrix elements $Q_{cc'}=\hbar b^\dagger_c b_{c'}$ as the overlap of the internal parts $b$ of the scattering wave functions in the incident channels $c$ and $c'$. We note that with such a factorised form, calculations of the moments of the Wigner time delay have some resemblance with those of the conductance and shot noise in the Landauer-B\"uttiker formalism of quantum transport \cite{bb00}. The statistics of the latter in chaotic cavities is determined by the Jacobi ensembles of RMT \cite{beenakker97}, the exact expressions for their first two cumulants being derived in \cite{ss06}. Transport moments of arbitrary order can be obtained most efficiently by exploiting the connection with the Selberg integral \cite{ss06,somm07a}, see further developments in \cite{novaes07,ssw08,novaes08,kss09} as well as \cite{vv08,ok08,ok09,liva11,mezz11,mezz12,mezz13,souz14} for other RMT studies on transport statistics. These results agreed with those from the semiclassical approach to quantum transport problems, as originally shown for the first two transport moments in \cite{heusleretal06,mulleretal07} and then generalized to an inverse channel expansion of all higher moments in \cite{bhn08,bk11}. Here, we provide for the first time a similar semiclassical justification of the Laguerre ensembles of RMT. In particular, we derive the exact expression \eref{eq:tWvar} for the variance of Wigner time delay semiclassically at any $M$ both for unitary and orthogonal symmetry (along with the other second moments of time-delays). We also extend previous work \cite{bk10,bk11} on establishing the universality for the moment generating functions.

In the next section, we formulate our starting point and develop a semiclassical approximation to the internal parts $b_c$ and establish diagrammatic sum rules. This representation is then applied in \sref{sec:secondmoms} to derive the second moments of time-delays at arbitrary $M$. Section~\ref{sec:momgenfuncs} deals with the moment generating function of the proper time-delays and works out the algorithmic approach for the first five terms in a $1/M$-expansion. Section~\ref{sec:concs} summarises our findings. Finally, we provide several Appendices with more technical details of our calculations, including sums for systems with TRS and a comparison to previous approaches, which we believe may be helpful for further development and applications of the method.

%%%%%%%%%%%%%%%%%%%%%%%%%%%%%%%%%%%%%%%%%%%%

\addtocontents{toc}{\vspace{-1em}}
\section{Resonance scattering approach}\label{sec:approach}

In the general scattering formalism, the resonance part of the scattering matrix can be represented in terms of the effective non-Hermitian Hamiltonian as follows \cite{soko89,verb85}:
\begin{equation} \label{eq:scatmatrmt}
S = 1 - \rmi V^{\dagger}\frac{1}{E-\Heff}V\,, \qquad \Heff=H-\frac{\rmi}{2}VV^{\dagger} \, .
\end{equation}
Here, the Hermitian $N \times N$ matrix $H$ corresponds to the Hamiltonian of the closed system, whereas the rectangular $N \times M$ matrix $V$ consists of the decay amplitudes that couple $N$ discrete energy levels to $M$ decay channels. These amplitudes are commonly treated as energy-independent quantities, which can be justified by considering resonance phenomena away from the open channel thresholds.
Substituting \eref{eq:scatmatrmt} into \eref{eq:Q}, one can readily find the representation of the Wigner-Smith matrix \cite{sz97},
\begin{equation} \label{eq:wsmatrmt}
Q = \hbar V^{\dagger} \frac{1}{(E-\Heff)^{\dagger}} \frac{1}{E-\Heff} V  \equiv \hbar b^{\dagger} b  \, ,
\end{equation}
in terms of the $N\times M$ matrix $b=(E-\Heff)^{-1}V$. In such an approach, the energy dependence enters only via the resolvent $(E-\Heff)^{-1}$ that describes the propagation in the open system governed by $\Heff$. The $N$-component vector $b_{c}$ (\ie the $c$th column of $b$) can therefore be treated \cite{sz97} as the intrinsic part of the scattering wave function initiated in the channel $c$ at energy $E$. The diagonal elements $q_c=Q_{cc}$ are then given by the norm of $b_c$, providing the interpretation as the average time delay of a wave packet in a given channel \cite{smith60}. Taking the sum over the diagonal elements, we find that the Wigner time delay is given by the total norm of the internal parts
\begin{equation} \label{eq:tWnorm}
 \tW = \frac{1}{M}\sum_{c=1}^{M} q_c \, , \qquad q_c = \hbar \|b_c\|^2 \, .
\end{equation}
This norm is also known as the dwell time, thus the two time characteristics coincide in the resonance approximation considered.

The factorized representation \eref{eq:wsmatrmt} already proved \cite{sss01,ss03} to be successful for deriving the exact (RMT) distributions of the proper time-delays (the eigenvalues of $Q$) in chaotic cavities. On the other hand, the \emph{partial} time-delays are defined by the energy derivative $t_c=\hbar\,\rmd{\phi_c}/\rmd{E}$ of the scattering eigenphases, $\phi_c$.  The exact distribution of the partial time-delays was found in \cite{fs97,fyod97a}.  Since the order of the diagonalisation and energy derivative is reversed for the proper and partial time-delays, they follow different statistics \cite{bfb97,bfb99}. In particular, for cavities with perfectly coupled leads (the case of interest here) the density of rates $t_c^{-1}$ reduces to $\chi_{M\beta}^2$ distribution characterised by the mean $\bar{t}=\tH/M=\bar\tW$ and the variance
\begin{equation}\label{eq:tcvar}
 \var(t_c) \equiv \frac{1}{\bar{t}^2}\left\langle \left(t_c-\bar{t}\right)^2 \right\rangle
  = \frac{2}{\beta M-2} \,.
\end{equation}
This should be compared with expression \eref{eq:tWvar} for $\var(\tW)$ that contains an extra factor $\frac{2}{M+1}$, thus reflecting the self-averaging property of the linear statistic \eref{eq:tWnorm} and diminishing correlations when $M$ grows.  It is worth noting that the covariance of two partial time-delays can also be computed exactly with the help of their joint distribution found in \cite{savi01}, with the explicit result being
\begin{equation}\label{eq:tccov}
 \mathrm{cov}(t_1,t_2) \equiv \frac{\langle t_1 t_2 \rangle}{\bar{t}^2}-1
  = \frac{2}{(M+1)(\beta M-2)} \,.
\end{equation}
More importantly, it was also shown in \cite{savi01} that at perfect coupling the statistical properties the partial time-delays become equivalent to those of the diagonal elements of the `symmetrised' Wigner-Smith matrix, $Q_s=S^{-1/2}QS^{1/2}$, introduced and studied in \cite{bfb97}. For the unitary case of systems without TRS, $S$ becomes statistically independent of $Q$ \cite{bfb97} and the symmetrisation process does not change the statistics of the diagonal elements so that $q_c$ also follow the same distribution, in particular, $\var(q_c)=\var(t_c)$. This is not the case for the other symmetry classes. However, traces of $Q$ and its powers are insensitive to such a symmetrisation, giving the identity
\begin{equation}\label{eq:twfromtc}
\var(\tW)=\frac{1}{M^2}\left[M\var(t_c)+M(M-1)\mathrm{cov}(t_1,t_2)\right] \, .
\end{equation}
Equations \eref{eq:tcvar} and \eref{eq:tccov} therefore readily yield the variance of the Wigner time delay in the form of \eref{eq:tWvar}. Likewise, we can arrive at the same result by using the variance \cite{mezz12} and covariance \cite{mart14} of the proper time-delays. For later use we quote the relevant expression for the second moment \cite{mezz11,mezz12}
\begin{equation}\label{eq:m2}
 m_2 \equiv \frac{1}{M} \langle\mathrm{Tr}(Q^2)\rangle = \frac{2\beta M^2\,\bar\tW^2}{(M+1)(\beta M-2)} \,.
\end{equation}

The semiclassical approximation for the internal parts $b_c$ developed below determines these time-delay quantities in terms of certain scattering trajectories and thus allows us to study the universality of RMT predictions for individual systems.

\subsection{The semiclassical approximation}

In this section we derive a semiclassical approximation for the Wigner time delay that is based on
representation \eref{eq:wsmatrmt} of the Wigner-Smith matrix $Q$. It is the third semiclassical approximation after \eref{eq:tWsemipos} and \eref{eq:tWsemi}. The starting point is the Green function $\mathcal{G}(\br,\br',E)$ of the open cavity with an arbitrary number of leads.
Its semiclassical approximation is a sum over all trajectories from $\br'$ to $\br$ \cite{gutzwiller71,gutzwiller90}
\begin{equation} \label{eq:greensemi}
\fl \mathcal{G}(\br,\br',E)
\approx \frac{1}{\rmi \hbar \sqrt{2 \pi \rmi \hbar}} \sum_\gamma \frac{1}{\sqrt{v_\gamma \, v_\gamma' \, | (M_{\gamma})_{12} |}}
\exp\left(\frac{\rmi}{\hbar} \mathcal{S}_\gamma - \frac{\rmi \, \pi}{2} \nu_\gamma \right) \, .
\end{equation}
Here, $\mathcal{S}_\gamma = \int_\gamma {\boldsymbol p} \, d {\boldsymbol q}$ is the action along the trajectory $\gamma$, $\nu_\gamma$ the number of conjugate points, and $v_\gamma'$ ($v_\gamma$) is the speed at initial (final) point (for a cavity without potential $v_\gamma' = v_\gamma$). Furthermore, $M_{\gamma}$ denotes the stability matrix that describes linearised motion near the trajectory. It connects perpendicular deviations from the trajectory at the end point to those at the initial point
\begin{equation} \label{eq:stabmat}
\fl\left( \begin{array}{c} d q_\perp \\ d p_\perp \end{array} \right) = M_\gamma \,
 \left( \begin{array}{c} d q_\perp' \\ d p_\perp' \end{array} \right) =
 \left( \begin{array}{cc} (M_\gamma)_{11} & (M_\gamma)_{12}
      \\ (M_\gamma)_{21} & (M_\gamma)_{22} \end{array} \right)
 \left( \begin{array}{c} d q_\perp' \\ d p_\perp' \end{array} \right)
\, .
\end{equation}
Formally, the effective Hamiltonian $\Heff$ of an open cavity is an infinite dimensional operator \cite{pich01,stoec02i,savi03}, corresponding to $N \rightarrow \infty$ of the matrix truncation in \eref{eq:scatmatrmt} and \eref{eq:wsmatrmt}. The position representation of the resolvent $(E-\Heff)^{-1}$  can then be identified with the Green function $\mathcal{G}(\br,\br',E)$, whereas $V$
corresponds to a projection onto the transverse wavefunctions in the leads such that \cite{fl81}
\begin{equation} \label{eq:GV}
\fl \langle  \br | \, b_c =
\langle  \br | \frac{1}{E - \Heff} V_c = \sqrt{\hbar v'_\parallel} \int_0^W d y' \, \mathcal{G}(\br,(x',y'),E) \, \Phi_{n(c)}(y') \, .
\end{equation}
The integration is over the cross section at the beginning of the lead that contains the $c$th incoming mode, $v'_\parallel= \frac{\hbar}{m} k_\parallel= \frac{\hbar}{m} \sqrt{k^2 - (n \pi/W)^2}$
is the longitudinal velocity ($m$ is the mass) and $W$ is the lead width. The corresponding transverse wavefunction is
\begin{equation} \label{eq:phin}
\Phi_n(y) = \sqrt{\frac{2}{W}} \sin \left( \frac{n \pi y}{W} \right) \, .
\end{equation}
Note that $1 \leq c \leq M$ labels the modes in all leads, whereas $n$ labels the modes in one particular lead, so the choice of the lead and $n$ depend on $c$.

The semiclassical approximation for the internal part $\langle \br | \, b_c$ follows by evaluating the integral in \eref{eq:GV} in stationary phase approximation. After writing the sine in \eref{eq:phin} as sum of two complex exponentials, the stationary phase condition reads
\begin{equation} \label{eq:statphascond}
\frac{\partial \mathcal{S}}{\partial y'} = -p_y' = - \frac{\bar{n} \hbar \pi}{W} \qquad \Longrightarrow \qquad \sin \theta_{\bar{n}} = \frac{ \bar{n} \pi}{k W} \, ,
\end{equation}
where $\bar{n} = \pm n$. This fixes the starting angle of the trajectories ($\sin \theta = p_y'/p'$) entering the cavity. Performing the stationary phase approximation results in
\begin{equation} \label{eq:bcsemi}
\langle  \br | \, b_c \approx \frac{1}{\sqrt{\hbar}} \sum_{\gamma(c \rightarrow \br)} A_\gamma \rme^{ \frac{\rmi}{\hbar} \mathcal{S}_\gamma } \, ,
\end{equation}
where the sum runs over all trajectories that enter the cavity with the angle fixed by \eref{eq:statphascond} and end at $\br$. The amplitudes are given by
\begin{equation} \label{eq:amplisemi3}
A_\gamma = \frac{- {\rm sign} (\bar{n}) }{\sqrt{2 v W \cos \theta_{\bar{n}} \, | (M_{\gamma})_{11} |}} \,
\exp \left(\frac{\rmi \bar{n} \pi y'}{W} - \frac{\rmi \, \pi}{2} \mu_\gamma \right) \, ,
\end{equation}
where $\mu_\gamma$ is the number of conjugate points for neighbouring trajectories with the same entrance angle.

The expressions \eref{eq:bcsemi} and \eref{eq:amplisemi3} allow us to represent the elements of the time-delay matrix \eref{eq:wsmatrmt} in terms of the trajectories specified above. In particular, the semiclassical approximation for the Wigner time delay follows from \eref{eq:tW} and \eref{eq:wsmatrmt} as
\begin{equation} \label{eq:tWsemi3}
\tW \approx \frac{1}{M} \sum_{c=1}^M \; \int \rmd^2 \br \sum_{\gamma, \gamma' (c \rightarrow \br)}
A_\gamma A_{\gamma'}^* \, \rme^{ \frac{\rmi}{\hbar} ( \mathcal{S}_\gamma - \mathcal{S}_{\gamma'}) } \, ,
\end{equation}
where the integral is over the interior of the cavity. This is the new representation for $\tW$ that serves as the starting point for the semiclassical calculations in this article.

We first apply \eref{eq:tWsemi3} to calculate the mean time delay. The approximation sums over pairs of trajectories that contribute with highly oscillatory terms. After spectral averaging most terms can be neglected and the only remaining terms are from pairs of trajectories that are correlated. These pairs will be discussed in the following.

The trajectories involved in \eref{eq:tWsemi3} are similar to those that occur in the semiclassics of the current density \cite{kuipersetal09} which in turn is related to the survival probability \cite{waltneretal08,gutierrezetal09}.

\subsection{Diagonal approximation for the mean time delay}

The leading contribution to the average time delay comes from the diagonal approximation where $\gamma' = \gamma$:
\begin{equation} \label{eq:tWdiag1}
\langle \tW \rangle \approx \left\langle \frac{1}{M} \sum_{c=1}^M \; \int \rmd^2 \br \sum_{\gamma (c \rightarrow \br)} |A_\gamma|^2 \right\rangle \, .
\end{equation}
The evaluation of this expression requires a sum rule for the type of trajectories in \eref{eq:tWdiag1}; see also \cite{sieber99} for related sum rules. To obtain the sum rule, we fix one of the leads and consider the probability density that trajectories starting in the opening with angle $\theta$ and energy $E$ will arrive after time $T$ at point $\br$,
\begin{equation} \label{eq:probdens}
P(\br,T,\theta,E)  = \frac{1}{W} \int_0^{W} \delta(\br(T) - \br) \, \rmd y' \, ,
\end{equation}
where $y'$ denotes the position in the opening, and the initial conditions of $\br(T)$ are determined by $y'$, $\theta$ and $E$. The integral can be evaluated in local coordinates that are parallel and perpendicular to the trajectory,
\begin{equation} \label{eq:probdens2}
 P_\varepsilon(\br,T,\theta,E) = \sum_\gamma \frac{1}{v W \cos \theta \, | (M_{\gamma})_{11} |} \delta_\varepsilon(T - T_\gamma) \, .
\end{equation}
This sum runs over all trajectories and has wild oscillations which can be damped by a conventional smoothing of the delta-function $\delta \rightarrow \delta_\varepsilon$.

In an open chaotic cavity with area $A$ the asymptotic form of the probability density is $P_\varepsilon(\br,T,\theta,E) \sim \rme^{- \mu T}/A$ as $T \rightarrow \infty$. Here the exponential term describes the asymptotic escape of trajectories and the $1/A$ reflects the fact that each end point in the cavity is equally likely.
We hence obtain the following sum rule
\begin{equation} \label{eq:sumrule}
\sum_{\gamma (c \rightarrow \br)} |A_\gamma|^2 \, \delta_\varepsilon (T - T_\gamma)
\sim \frac{1}{A} \rme^{- \mu T} \, .
\end{equation}
Note that channel $c$ corresponds to two angles $\theta$ which cancels a factor $1/2$ coming from \eref{eq:amplisemi3}. With this sum rule we can evaluate the diagonal approximation and obtain
\begin{equation} \label{eq:tWdiag2}
\langle \tW \rangle \approx \frac{1}{M} \sum_{c=1}^M \; \int \rmd^2 \br \, \frac{1}{A} \int \rme^{- \mu T} \rmd T = \frac{1}{\mu} \, .
\end{equation}
This is already the correct expression for the mean time delay. We will now show that off-diagonal contributions leave this result intact.

\subsection{First off-diagonal corrections for the mean time delay} \label{sec:leading off}

Off-diagonal contributions come from trajectories that have close \emph{self-encounters} in which two or more stretches of the trajectory are almost parallel or anti-parallel \cite{sr01,mulleretal04,mulleretal05,rs02,heusleretal06,mulleretal07}.
These trajectories have close neighbours that differ in the way in which
the remaining longer parts of the trajectory, the \emph{links}, are connected in the encounter regions.

\begin{figure}
 \centering
\includegraphics[width=\textwidth]{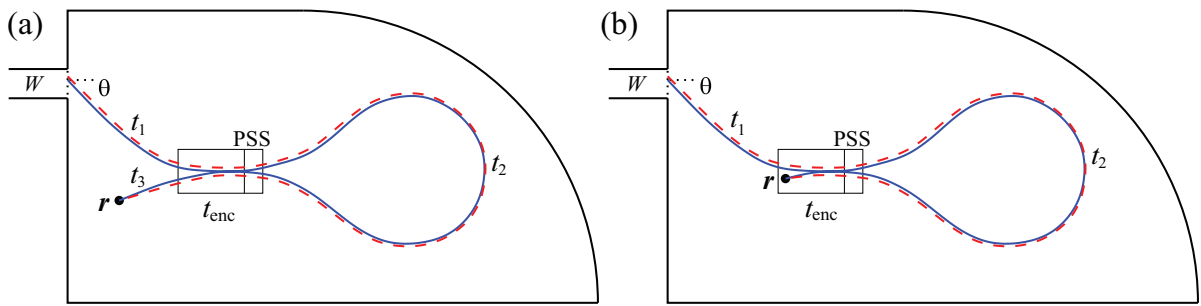}
 \caption{(a) Schematic picture of a trajectory with a self-encounter (full line). The neighbouring trajectory (dashed line) traverses the loop in the opposite direction. The encounter region is indicated by a rectangular box and contains a Poincar\'e surface of section (PSS). (b) The so-called \emph{one-leg loop} corresponds to trajectories that have their end point in the encounter region, yielding a semiclassical contribution of the same order as in (a).}
 \label{fig:leadingoff}
\end{figure}

The simplest example is a trajectory with one encounter as shown in \fref{fig:leadingoff}(a). The neighbouring trajectory (dashed line) starts with the same angle $\theta$ and arrives at the same end point $\br$, but it traverses the loop in the opposite direction. For this reason these pairs exist only in systems with TRS. For the evaluation of the semiclassical contribution of these orbits we follow \cite{heusleretal06,mulleretal07}.

The encounter is described in a Poincar\'e surface of section in the encounter region. The relative distance of the two piercings of the original trajectory through the Poincar\'e surface is specified by coordinates $s$ and $u$ along the stable and unstable manifolds. This information is sufficient to determine the neighbouring trajectory and the action difference that is given by $\mathcal{S}_{\gamma} - \mathcal{S}_{\gamma'} = su$ in the linearised approximation.
The duration of the encounter region is specified by requiring that the distance of the two stretches of the trajectory along the stable and unstable directions remain smaller than some small constant $c$, leading to
\begin{equation} \label{eq:encdur1}
t_{\rm enc}(s,u) = \frac{1}{\lambda} \ln \frac{c^2}{|su|} \, ,
\end{equation}
where $\lambda$ is the Lyapunov exponent.
The summation over the trajectory pairs is done by applying probabilistic arguments. Let $w_T(s,u)$ be the probability density in a chaotic system that a trajectory of long duration $T$ has a close self-encounter that is specified by coordinates $s$ and $u$.
The contribution of the trajectory pairs to the average time delay can then be expressed as
\begin{equation}
\langle \tW^{1a} \rangle = \frac{1}{M} \sum_{c=1}^M \; \int \rmd^2 \br \int \rmd s \, \rmd u \sum_{\gamma, \gamma' (c \rightarrow \br)}
|A_\gamma|^2 \, w_T(s,u) \, \rme^{ \frac{\rmi}{\hbar} s u } \, .
\end{equation}
The density $w_T(s,u)$ is given by an integral over the first two link durations
\begin{equation}
w_T(s,u) = \int_0^{T - 2 t_{\rm enc}} \rmd t_1 \int_0^{T - 2 t_{\rm enc} - t_1} \rmd t_2 \; \frac{1}{\Omega \, t_{\rm enc}(s,u)} \, ,
\end{equation}
where $\Omega$ is the volume of the surface of constant energy in phase space. Applying the sum rule \eref{eq:sumrule} and
changing the integral over the orbit length $T$ to an integral over the third link duration results in
\begin{equation} \label{eq:tW1a2} \fl
\langle \tW^{1a} \rangle = \frac{1}{M} \sum_{c=1}^M \; \int \rmd^2 \br \frac{1}{A} \int_0^\infty \rmd t_1 \, \rmd t_2 \, \rmd t_3
\int  \rmd s \, \rmd u \, \frac{\rme^{ \frac{\rmi}{\hbar} s u } \, \rme^{-\mu(t_1+t_2+t_3+t_{\rm enc}(s,u))}}{\Omega \, t_{\rm enc}(s,u)}  \, ,
\end{equation}
Equation \eref{eq:tW1a2} contains a small correction to the sum rule \eref{eq:sumrule} that is necessary when applied to trajectories with self-encounters. Namely, the trajectory time $T$ in the exponent has to be replaced by the \emph{exposure} time that counts each encounter duration only once. The reason is that if a trajectory does not escape during the first traversal of an encounter region, it will not escape during subsequent traversals of this region.

The integral over $s$ and $u$ can now be evaluated by noting that the only contribution that survives in the semiclassical limit is one where the encounter duration $t_{\rm enc}$ in the denominator is exactly cancelled by an encounter duration in the numerator. In other words, after expanding  $\exp(-\mu t_{\rm enc})$ in a Taylor series the only term that survives is the linear term in $t_{\rm enc}$. With $\int \rmd s \, \rmd u \exp(\rmi s u / \hbar) = 2 \pi \hbar$, we finally obtain
\begin{equation} \label{eq:tW1a3}
\langle \tW^{1a} \rangle = \left( \frac{1}{\mu} \right)^3 \, \left( - \frac{\mu}{\tH} \right) =  - \frac{\tH}{M^2} \, ,
\end{equation}
where we have used $\tH = \Omega/2 \pi \hbar$ and $\mu = M/\tH$.

The contribution \eref{eq:tW1a3} would lead to a deviation from the correct result \eref{eq:tWdiag1}. There is, however, a further contribution of the same order. It arises from the so-called \emph{one-leg loops}, which are correlated trajectories that have an end point in an encounter region as in \fref{fig:leadingoff}(b). These type of correlations do not occur in transport problems, but they arise when trajectories have one or both their end points in the cavity as for example in problems involving the survival probability, the current density or the fidelity \cite{waltneretal08,gutierrezetal09,kuipersetal09,gutkinetal10}.

The encounter regions of one-leg loops require a different treatment than the usual encounters. It can be shown, however, that this difference can effectively be taken into account by adding a further factor of $t_{\rm enc}$ in the integral  over $s$ and $u$ \cite{kuipersetal09}. So the contribution of the one-leg loop trajectories differs from \eref{eq:tW1a3} by a missing integration over the third link time $t_3$ and an additional factor of $t_{\rm enc}$,
\begin{equation} \label{eq:tW1b2} \fl
\langle \tW^{1b} \rangle = \frac{1}{M} \sum_{c=1}^M \; \int \rmd^2 \br \frac{1}{A} \int_0^\infty \rmd t_1 \, \rmd t_2
\int  \rmd s \, \rmd u \, \frac{\rme^{ \frac{\rmi}{\hbar} s u } \, \rme^{-\mu(t_1+t_2+t_{\rm enc}(s,u))}}{\Omega}  \,.
\end{equation}
The integrals are then evaluated similarly as before and result in
\begin{equation} \label{eq:tW1b3}
\langle \tW^{1b} \rangle = \left( \frac{1}{\mu} \right)^2 \, \left( \frac{1}{\tH} \right) =  \frac{\tH}{M^2} \, .
\end{equation}
This contribution cancels exactly the contribution \eref{eq:tW1a3}.

One could further consider shrinking the first link in \fref{fig:leadingoff}(b) so that the encounter moves into the lead creating a `coherent back-scattering' type of diagram.  However the freedom of how much of the encounter box overlaps with the lead provides a further factor of the encounter time.  In calculating the semiclassical contribution, the integrand then becomes at least linear in $t_{\rm enc}$ and the integral vanishes in the semiclassical limit.  In general our attention may simply be restricted to diagrams with at most one end point in each encounter \cite{waltneretal08,gutierrezetal09,kuipersetal09,gutkinetal10}.

\subsection{Higher off-diagonal corrections for the mean time delay}

For higher-order corrections one considers trajectories with arbitrarily many self-encounters. These self-encounters can involve two or more stretches of a trajectory that are almost parallel or anti-parallel, where the latter case requires TRS. One speaks of an $l$-encounter if it involves $l$ stretches of a trajectory. The types of a trajectory's encounters are detailed in a vector $\v$ whose $l$th component $v_l$ specifies the number of $l$-encounters. The total number of encounters is thus $V = \sum_l v_l$, and the total number of stretches in all encounter regions is $L = \sum_l l \, v_l$.

Trajectories with self-encounters have close neighbouring trajectories that differ in the way in which the links are connected in the encounter regions. For a given vector $\v$ there are many different configurations in which the encounters and the reconnections can be arranged along a trajectory pair.
The number of these \emph{structures} or \emph{families} is denoted by $\mathcal{N}(\v)$. The action difference of a trajectory pair can again be determined in terms of the separation of the trajectory stretches in the encounter regions.
There are now altogether $L-V$ pairs of coordinates in the stable and unstable directions ($l-1$ pairs for each $l$-encounter). These coordinates are combined into vectors $\s$ and $\bu$, and in the linearised approximation one has $\mathcal{S}_{\gamma} - \mathcal{S}_{\gamma'} = \s \bu$.

The definition of the encounter duration \eref{eq:encdur1} is generalised to arbitrary $l$-encounters by requiring that the separations of the $l$ trajectory stretches remain smaller than some constant $c$ in the stable and unstable directions,
\begin{equation} \label{eq:endcur2}
t_{\rm enc}^\alpha(\s,\bu) = \frac{1}{\lambda} \ln \frac{c^2}{\max_j |s_{\alpha,j}| \times \max_j |u_{\alpha,j}|} \, ,
\end{equation}
where $\alpha$ labels the encounters, $1 \leq \alpha \leq V$. One applies again probabilistic arguments to replace the summation over trajectory pairs in \eref{eq:tWsemi3} by one over self-encounters,
\begin{equation}
\fl \langle \tW^{\v a} \rangle = \frac{\mathcal{N}(\v)}{M} \sum_{c=1}^M \; \int \rmd^2 \br \int \rmd \s \, \rmd \bu \sum_{\gamma, \gamma' (c \rightarrow \br)}
|A_\gamma|^2 \, w_T(\s,\bu) \, \rme^{ \frac{\rmi}{\hbar} \s \bu } \, ,
\end{equation}
where $w_T(\s,\bu)$ is the probability density that a trajectory of long duration $T$ has self-encounters that are specified by the vector $\v$ and the separations $\s$ and $\bu$. This density can be expressed by an integral over the first $L$ link durations
\begin{equation} \fl
w_T(\s,\bu) = \int_0^{T - t_{\rm enc}} \rmd t_1 \ldots \int_0^{T - t_{\rm enc} - t_1 - \ldots - t_{L-1}} \rmd t_L \; \;
\frac{1}{\Omega^{L-V} \, \prod_\alpha t_{\rm enc}^{\alpha}(\s,\bu)} \, .
\end{equation}
At a final step, the sum rule \eref{eq:sumrule} is applied. As earlier in \eref{eq:tW1a2}, it requires replacing the time in the exponent by the exposure time counting all encounter times only once. After replacing the integral over the trajectory time $T$ by an integral over the final
link duration, one obtains
\begin{equation} \label{eq:tWva2} \fl
\langle \tW^{\v a} \rangle = \frac{\mathcal{N}(\v)}{M} \sum_{c=1}^M \; \int \rmd^2 \br \frac{1}{A} \; \; \left( \int_0^\infty \rmd t \,  \rme^{-\mu t} \right)^{L+1}
\int  \rmd \s \, \rmd \bu \, \frac{\rme^{ \frac{\rmi}{\hbar} \s \bu } \, \rme^{-\mu \sum_\alpha t_{\rm enc}^{\alpha}(\s,\bu)
}}{\Omega^{L-V} \, \prod_\alpha t_{\rm enc}^{\alpha}(\s,\bu)}  \, .
\end{equation}
In the integrals over the $\s$ and $\bu$ coordinates the only terms that survive the semiclassical limit are those where all the encounter times in the denominator are exactly cancelled by corresponding encounter times in the numerator. The leading contribution thus again arises from the linear terms in the Taylor expansion of $\exp(-\mu \sum_\alpha t_{\rm enc}^{\alpha}(\s,\bu))$.
Using $\int \rmd \s \, \rmd \bu \exp(\rmi \s \bu / \hbar) = 2 \pi \hbar$, the final result reads
\begin{equation} \label{eq:tWva3}
\langle \tW^{\v a} \rangle = \mathcal{N}(\v) \; \; \frac{1}{\mu^{L+1}}  \, \frac{(-\mu)^V}{\tH^{L-V}} =   \mathcal{N}(\v) \, (-1)^V \, \frac{\tH}{M^{L-V+1}} \,.
\end{equation}

As in the transport problem \cite{heusleretal06,mulleretal07}, one can identify simple diagrammatic rules from this result. Each link contributes by a factor $1/M$, and each encounter contributes a factor $(-M)$.  The Heisenberg times cancel up to one. Note further that the sum over the channels gives a factor of $M$ that cancels the prefactor $1/M$ in \eref{eq:tW}.

As we have seen in section \ref{sec:leading off}, there are additional trajectory correlations for the new semiclassical representation \eref{eq:tWsemi3} of the Wigner time delay due to the one-leg loops that do not occur in the transport problem. In fact, for every trajectory configuration or \emph{structure} in the above calculation there is a corresponding configuration where the end point is now inside the last encounter. The semiclassical calculation can be easily modified to obtain these additional contributions. The difference to \eref{eq:tWva2} is that the integral over the final link duration is missing, and the integrals over the $\s$ and $\bu$ coordinates contain an additional factor of the last encounter time \cite{kuipersetal09}. These modifications lead to
\begin{equation} \label{eq:tWvb3}
\langle \tW^{\v b} \rangle = \mu \left(-\frac{1}{\mu} \right) \langle \tW^{\v a} \rangle = - \langle \tW^{\v a} \rangle \, .
\end{equation}
As expected all off-diagonal terms cancel. This calculation allows us to extend the diagrammatic rules: Encounters that include an end point simply contribute a factor of~$1$.

\subsection{Diagrammatic rules for higher moments} \label{sec:diagrules}

The previous section has shown one main advantage of the new semiclassical approach. It results in the simple diagrammatic rules that are similar to the ones in transport problems \cite{mulleretal07}, thus strongly simplifying the calculations. As in transport, one can generalise the diagrammatic rules to higher moments. We sketch this by considering the moments of the proper time-delays defined as
\begin{equation} \label{eq:momentsdef}
m_n=\frac{1}{M} \left\langle \Tr\left[Q^n\right] \right\rangle \,.
\end{equation}
The representation \eref{eq:wsmatrmt} and the semiclassical approximation \eref{eq:bcsemi} lead to
\begin{equation} \label{eq:momentssemi}  \fl
m_n = \frac{1}{M} \left\langle \sum_{i_1,\ldots,i_n=1}^M \int \rmd^2 \br_1 \ldots \rmd^2 \br_n \sum_{{\boldsymbol \gamma},{\boldsymbol \gamma'}}
\left( \prod_{j=1}^n A_{\gamma_j} A_{\gamma'_j}^* \right)
\rme^{\rmi(\mathcal{S}_{\boldsymbol \gamma} - \mathcal{S}_{{\boldsymbol \gamma'}})/\hbar} \right\rangle \, .
\end{equation}
Here $i_1, \ldots , i_n$ label the $n$ incoming channels and $\br_1, \ldots, \br_n$ denote the $n$ final points. The symbols
${\boldsymbol \gamma} = \{\gamma_1, \ldots , \gamma_n\}$ and  ${\boldsymbol \gamma'} = \{\gamma'_1, \ldots , \gamma'_n\}$ stand
for two sets of $n$ trajectories, where $\gamma_j$ goes from channel $i_j$ to the point $\br_j$, and $\gamma'_j$ from $i_{j+1}$ to $\br_j$ (we identify $i_{n+1}$ with $i_1$).
The total actions of the two sets are $\mathcal{S}_{\boldsymbol \gamma} = \sum_j \mathcal{S}_{\gamma_j}$ and
$\mathcal{S}_{\boldsymbol \gamma'} = \sum_j \mathcal{S}_{\gamma'_j}$, respectively.

Dealing with the correlated trajectories that survive the spectral averaging in \eref{eq:momentssemi} now involves two trajectory sets, ${\boldsymbol \gamma}$ and ${\boldsymbol \gamma'}$. The trajectories in the set ${\boldsymbol \gamma}$ have encounters in which two or more trajectory stretches are almost parallel or anti-parallel. The set ${\boldsymbol \gamma'}$ follows the set ${\boldsymbol \gamma}$ very closely along the links, but differs in the way those are connected in the encounter regions. One can then replace the double sum over ${\boldsymbol \gamma}$ and ${\boldsymbol \gamma'}$ by a single sum over ${\boldsymbol \gamma}$ plus an integral over a probability density for the self-encounters. The result can be split into contributions from links and encounters according to the following diagrammatic rules:
\begin{itemize}
 \item The summation over each incoming channel gives a factor of $M$.
 \item Each link contributes a factor of $1/M$.
 \item Each encounter gives a factor of $(-M)$, unless it contains an end point.
 \item An encounters contributes a factor of $1$ if it contains one end point, and a factor of $0$ if it contains more than one end point.
\end{itemize}
There is furthermore an overall factor of $T_H^n$, and a factor of $1/M$ from \eref{eq:momentsdef}. The rules for the encounters follow from the fact that each end point inside an encounter provides an additional factor of the encounter time.

\begin{figure}
  \centering
  \includegraphics{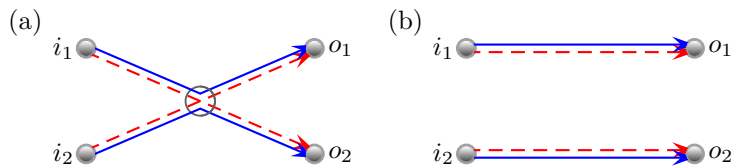}
  \caption{(a) A quadruplet with a single encounter. (b) A quadruplet involving independent links.}
  \label{fig:leadingorderquads}
\end{figure}

As an example, we discuss the leading order contribution to the second moment $m_2$.
In analogy to transport problems we denote the $j$th end point by $o_j$ instead of $\br_j$.
The simplest trajectory configuration with an encounter is the one that is schematically shown in \fref{fig:leadingorderquads}(a). The two trajectories belonging to ${\boldsymbol \gamma}$ are shown by the full lines and go from channel $i_1$ to end point $o_1$
and from $i_2$ to $o_2$, respectively. They have one encounter that is indicated by the circle. The neighbouring trajectories belonging to ${\boldsymbol \gamma'}$ (dashed lines) go from $i_1$ to $o_2$ and from $i_2$ to $o_1$, respectively.
According to the above rules we obtain the following contribution of this configuration to $m_2$
\begin{equation} \label{eq:xm2}
\frac{-M^3}{M^4} \, \frac{T_H^2}{M} = - \frac{1}{\mu^2} \, .
\end{equation}
This contains $(-M)$ from the encounter, $M^2$ from the incoming channels, $1/M^4$ from the links, and the overall factor $T^2_H/M$.

\begin{figure}
  \centering
  \includegraphics{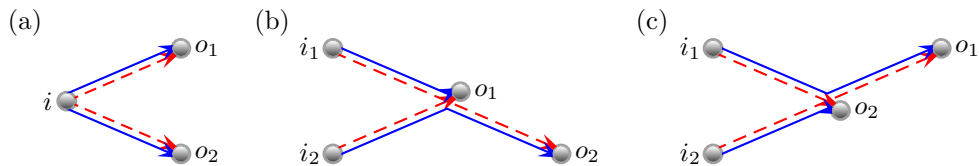}
  \caption{Three trajectory configurations that contribute to the leading order of the second moment $m_2$. In (a) the incoming channels coincide,
  and (b) and (c) have one end point in the encounter. All three are limiting cases of \fref{fig:leadingorderquads}(a).}
  \label{fig:secondmoments}
\end{figure}

The simpler configuration in \fref{fig:leadingorderquads}(b) does not usually contribute to $m_2$ since the trajectories of ${\boldsymbol \gamma'}$
(dashed lines) don't connect the correct initial and final points. Note, however, that this configuration is possible if the incoming channels coincide.
Then one obtains the configuration in \fref{fig:secondmoments}(a). It contributes at the same order as \eref{eq:xm2} since it has two links, one encounter
and one incoming channel less. There is the further possibility that one of the end points is in the encounter as in \frefs{fig:secondmoments}(b)
and (c), which again contribute at the same order. All in all the result is
\begin{equation} \label{secondmomentleadingorder}
\left( \frac{-M^3}{M^4} + \frac{M}{M^2} + 2\frac{M^2}{M^3} \right) \, \frac{T_H^2}{M} = 2 \frac{T_H^2}{M^2} = \frac{2}{\mu^2}  \, .
\end{equation}
This is indeed the leading order term of $m_2$ in systems with or without TRS.

We have obtained it by considering \frefs{fig:leadingorderquads}(a) and \ref{fig:secondmoments}(a)-(c) as different trajectory configurations that all contribute to the moment. There is even simpler and alternative point of view in which one considers all the contributions in \eref{secondmomentleadingorder} to come from \fref{fig:leadingorderquads}(a). The diagrams in \frefs{fig:secondmoments}(a)-(c) are considered to be limiting cases of \fref{fig:leadingorderquads}(a) where either the encounter moves into the incoming lead or one of the end points moves into the encounter. These limiting cases can be included by changing the contribution of the encounter.

For the description we adopt the language from transport and call the initial points of ${\boldsymbol \gamma}$ $i$-leaves and the final points $o$-leaves. For the calculation of the moments $m_n$ we then have to take into account the following cases:
\begin{itemize}
 \item Encounters of size $l$ can move into the incoming lead when connected directly to $l$ $i$-leaves.
 \item The $o$-leaves can be moved into the encounter they are connected to, but each encounter can only take one at a time.
\end{itemize}
In terms of the semiclassical contributions, in both of these situations we change the rule for the affected encounter to include these cases. In the first case we lose $l$ links, $(l-1)$ channel summations and the $(-M)$ from the encounter, and in the second case we lose one link and the $(-M)$ from the encounter. The contribution of encounter $\alpha$ is hence changed to $M(-1+\delta+s_\alpha)$ with $\delta$ being 1 when the encounter can move into the lead (and 0 otherwise) and $s_\alpha$ the number of $o$-leaves attached.

In our example, if we change the encounter contribution for \fref{fig:leadingorderquads}(a) in \eref{eq:xm2} from $(-M)$ to $M(-1+1+2)=2M$ then we obtain the same result as in \eref{secondmomentleadingorder}. One can also easily see
that the off-diagonal contributions for the average time delay all vanish, because the last encounter must have $\delta=0$ ($l$ is at least 2 and there is only 1 $i$-leaf) and $s_v=1$ since it is connected to the outgoing channel. The product of contributions is then 0.

We will show below that these rules can be employed systematically to calculate the second moments of the proper time-delays and the variance of the Wigner time delay, and they can as well be incorporated into the graphical framework
\cite{bhn08,bk10,kuipersetal11,bk11,bk12,bk13a,bk13b} to give moment generating functions.
Previous approaches to the time delay involved including an energy dependence so that calculating the semiclassical contribution of any diagram required knowledge of its complete structure. Here instead the semiclassical contributions are much closer to standard transport moments where only the difference in the number of links and encounters matters. The only additional complication is that now some information about the position of the $o$-leaves is necessary. Determining this is however much less demanding than treating full energy dependence.

%%%%%%%%%%%%%%%%%%%%%%%%%%%%%%%%%%%%%%%%%%%%

\addtocontents{toc}{\vspace{-1em}}
\section{Second moments} \label{sec:secondmoms}

The diagrammatic rules turn semiclassical calculations into a combinatorial problem of counting
trajectory configurations (also called \emph{structures} or \emph{diagrams}). Before calculating the second moments of various time-delay quantities, we recall that the structures that are relevant for transport moments \cite{heusleretal06,braunetal06,mulleretal07} (and hence also for the time delay) can be related to the structures of periodic orbit pairs that contribute to the two-point correlators in spectral statistics \cite{sr01,mulleretal04,mulleretal05}.

Correlated periodic orbit pairs are also described by a vector $\v$ whose elements $v_l$ count the number of $l$-encounters of the orbits. The total number of encounters is $V=\sum_l v_l$ while the number of links is $L=\sum_l lv_l$. The number of periodic orbit structures with a vector $\v$ and a labelled first link is $N(\v)$. These numbers can be determined recursively \cite{mulleretal05}.

In the following we deal with systems without TRS. The case of preserved TRS is briefly discussed in section \ref{sec:trs}, being mostly deferred to the Appendices.

\subsection{Counting diagrams in transport problems}

If a periodic orbit pair is cut along one of its links, it can be deformed into a pair of scattering trajectories which contributes to the first transport moment, the average conductance \cite{heusleretal06}. The cut link becomes two so that conductance diagrams have $L+1$ links. The number of structures $N(\v)$ of the scattering trajectories is the same as the one for periodic orbits. Since encounters contribute with a minus sign, terms in a $M^{-1}$ expansion of the conductance can be related to the sum \cite{heusleretal06}
\begin{equation} \label{unitCsum}
C_K=\sum_{\v}^{L-V=K}(-1)^V N(\v) \, , \qquad K \geq 1 \, .
\end{equation}
This is the relevant quantity to evaluate. Without TRS, one can show $C_K=0$ for $K\geq1$ so only the diagonal pair is important \cite{heusleretal06}.
It can be included in the formalism by defining $L=V=0$ and $C_0=1$ for it.

For the second transport moments (the shot noise and the conductance variance), the semiclassical diagrams involve four trajectories.
These trajectory quadruplets can be divided into two groups: $d$-quadruplets and $x$-quadruplets.
Consider a pair of trajectories $\gamma_1$ and $\gamma_2$ connecting channel $i_1$ to $o_1$ and $i_2$ to $o_2$ respectively. If the partner trajectory $\gamma_1'$ also connects channel $i_1$ to $o_1$ then we have a $d$-quadruplet. Otherwise if $\gamma_1'$ connects $i_1$ to $o_2$ we have an $x$-quadruplet.
The remaining partner trajectory $\gamma_2'$ must connect $i_2$ to the other outgoing channel.
For example, the diagram in \fref{fig:leadingorderquads}(a) is an $x$-quadruplet of trajectories meeting at a single 2-encounter while the diagram in \fref{fig:leadingorderquads}(b) is the simplest $d$-quadruplet. It is made up of two independent links.

Denote the number of $x$-quadruplet diagrams corresponding to a vector $\v$ by $N_x(\v)$, and the number of $d$-quadruplet diagrams by $N_d(\v)$,
then the conductance variance and the shot noise can be related to the sums
\begin{equation} \label{Dsum}
D_K=\sum_{\v}^{L-V=K}(-1)^V N_d(\v) \, ,
\end{equation}
\begin{equation} \label{Xsum}
X_K=\sum_{\v}^{L-V=K}(-1)^V N_x(\v) \, .
\end{equation}

These sums can again be related to periodic orbit pairs. Without TRS, as detailed in \cite{mulleretal07}, cutting periodic orbit pairs
twice along links leads to a $d$-type quadruplet. In general this can be done in $L(L+1)$ ways since we may cut the same link twice (and the order matters).
Note that the final $d$-quadruplet will have $L+2$ links. On the other hand, an $x$-quadruplet can be created by cutting out an entire 2-encounter from a periodic orbit pair.
In general this can be done in $2v_2$ ways and the final $x$-quadruplet will have one 2-encounter fewer and the same number of links as the periodic orbit pair.

With TRS, the cutting is more complicated, but in both symmetry cases the quantities $D_K$ and $X_K$ can be related to the following sums
\begin{equation} \label{Asum}
A_K=\sum_{\v}^{L-V=K}(-1)^V(L(\v)+1) N(\v) \, ,
\end{equation}
\begin{equation} \label{Bsum}
B_K=\sum_{\v}^{L-V=K}(-1)^V\frac{2v_2}{L(\v)} N(\v) \, .
\end{equation}
Without TRS, both are equal to 1 for even $K$ and 0 otherwise; and we have $D_K=A_K$ while $X_K=-B_{K+1}$ \cite{mulleretal07}.
We further have $D_0=1$ and $X_0=0$.

\subsection{Grouping diagrams}

The sums $X_K$ are known, but for the time delays we need to make a further distinction because of the different rules for the $o$-leaves.
We divide the $x$-quadruplets into two groups: we denote those where both final points are connected to the same encounter by $\tilde{x}$,
and those where they are connected to different encounters by $x'$.
For each vector $\v$ we count the structures in each group as $N_{\tilde{x}}(\v)$ and $N_{x'}(\v)$, with
\begin{equation}
N_x(\v) =  N_{\tilde{x}}(\v) + N_{x'}(\v) \, ,
\end{equation}
and we define the sums
\begin{equation} \label{Xtsum}
\tilde{X}_K=\sum_{\v}^{L-V=K}(-1)^V N_{\tilde{x}}(\v) \, ,
\end{equation}
\begin{equation} \label{Xpsum}
{X}'_K=\sum_{\v}^{L-V=K}(-1)^V N_{x'}(\v) = X_K-\tilde{X}_K \, .
\end{equation}
Only the $\tilde{X}_K$ part will contribute to the second moment of the time delays. This follows from the rules in section \ref{sec:diagrules}. If the two final points
are connected to different encounters then at least one of these encounters cannot be moved into the incoming lead and contributes with a factor of
$M(-1+\delta + s_\alpha)=0$.

We likewise partition the $d$-quadruplets and define the sums
\begin{equation} \label{Dtsum}
\tilde{D}_K=\sum_{\v}^{L-V=K}(-1)^V N_{\tilde{d}}(\v) \, ,
\end{equation}
\begin{equation} \label{Dpsum}
{D}'_K=\sum_{\v}^{L-V=K}(-1)^V N_{d'}(\v) \, ,
\end{equation}
where again we are only interested in the $\tilde{D}_K$ part.  However, along with $N_{\tilde{d}}(\v)$ and $N_{d'}(\v)$, we also need to take into
account a third group where one or both end points are not connected to an encounter at all.
These diagrams involve a direct link between an incoming channel and an end point, while the other trajectory pair can be any arbitrary conductance
(first moment) diagram. We then have
\begin{equation} \label{Drel}
D_K= \tilde{D}_K + {D}'_K + 2C_K - \delta_{K,0} \, ,
\end{equation}
where the last term is a correction to avoid overcounting the diagram in \fref{fig:leadingorderquads}(b) made up of two direct links.

\subsection{Manipulating diagrams in systems without TRS}\label{sec:unitarymanipdiag}

\begin{figure}
  \centering
  \includegraphics{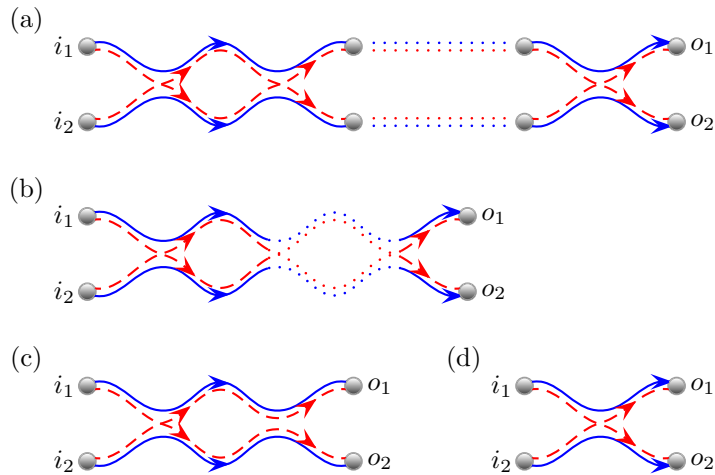}
  \caption{The final links of any $\tilde{d}$-quadruplet leave from the same $l$-encounter, with $l=2$ for the example in (a).  Appending a 2-encounter creates an $x$-quadruplet and we start to shrink both intervening links to arrive at (b).  Shrinking the links further until the two final encounters merge gives figure (c) which is identical to the $x$-quadruplet in (d) or \fref{fig:leadingorderquads}(a).  Reversing the process, we may add a 2-encounter to the final two links of any $x$-quadruplet to obtain a $\tilde{d}$-quadruplet ending in a 2-encounter.}
  \label{fig:dshrink1}
\end{figure}

\begin{figure}
  \centering
  \includegraphics{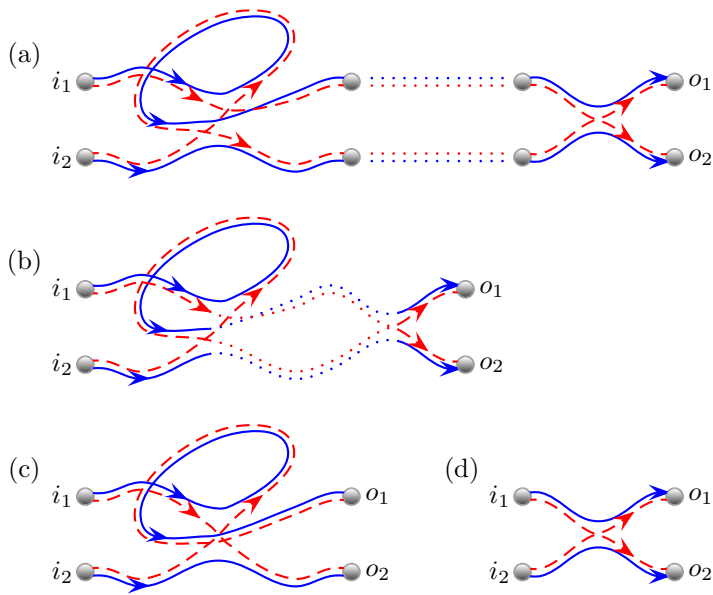}
  \caption{Here the final links of a $\tilde{d}$-quadruplet leave from the same 3-encounter to which we append a new 2-encounter in (a). Shrinking both intervening links we pass through (b) to end up at (c).  The link ending in the end point $o_1$ is now no longer involved in the encounter and can be unwound to create the $x$-quadruplet in (d) or \fref{fig:leadingorderquads}(a).  Reversing the process, we may merge either of the final two links of any $x$-quadruplet into the last $l$-encounter connected to the other end point and create $\tilde{d}$-quadruplet ending in a larger $(l+1)$-encounter.}
  \label{fig:dshrink2}
\end{figure}

To obtain some information about $\tilde{D}_K$ and $\tilde{X}_K$ we need to build a recursion relation between them. We start by considering ways in which an
$\tilde{d}$-quadruplet can be obtained from other diagrams. For a $\tilde{d}$-quadruplet both end points are connected to the same encounter (of size $l$ say), and we can add a 2-encounter at the end. Now we have an $x$-quadruplet whose final 2-encounter has two links connected to the same $l$-encounter. Next we shrink the connecting links and merge the two encounters. If $l=2$ both 2-encounters vanish and we have an arbitrary $x$-quadruplet with one encounter and 2 links fewer than the original $\tilde{d}$-quadruplet.  This process is depicted in \fref{fig:dshrink1}. If $l>2$ there are two possibilities: A link could separate from the encounter which becomes one smaller ($l \rightarrow l-1$) to leave an arbitrary $x$-quadruplet with the same number of encounters but one link fewer than before. An example is given in \fref{fig:dshrink2}. Alternatively the encounter could break into two separate encounters each connected to an outgoing channel hence giving a $x'$-quadruplet. This has one more encounter than the original $\tilde{d}$-quadruplet and the same number of links.

Reversing the three processes, we can obtain any $\tilde{d}$-quadruplet in exactly one of the following ways. We may add a 2-encounter to an arbitrary $x$-quadruplet (\fref{fig:dshrink1}). Alternatively for any $x$-quadruplet, we may join the link before an end point into the last encounter before the other end point (\fref{fig:dshrink2}). We may also join the final distinct encounters of any $x'$-quadruplet by pulling out a 2-encounter to create a $\tilde{d}$-quadruplet. Accounting for the minus sign from each encounter and the change in $L$ and $V$ we arrive at the relation
\begin{eqnarray} \label{DXrel}
\tilde{D}_K &=& - X_{K-1} + 2X_{K-1} - X'_{K-1} \nonumber \\
&=& \tilde{X}_{K-1}  \, ,
\end{eqnarray}
since $X'_K=X_K-\tilde{X}_K$. Another way to look at this is the following: one can describe an $l$-encounter as a cyclic permutation $(a_1\ldots a_l)$ where the $a_j$ are labels corresponding to the order encounter stretches are visited in the entire diagram \cite{mulleretal05}. If stretches $a_i$ and $a_j$ are connected to the outgoing channel, then adding a 2-encounter and shrinking the intervening links corresponds to multiplying (on the left) by $(a_i \, a_j)$. This breaks the $l$-cycle into a $k$ and $l-k$ cycle.  If $k$ or $l-k$ are 1 then a link is separated from the encounter (leaving only a link if $l=2$) otherwise the encounter breaks into two.

We can repeat the same process of adding a 2-encounter and shrinking the adjoining links but starting with an $\tilde{x}$-quadruplet. This again leads to three cases of $d$-quadruplets with a lower value of $L-V$: one with a 2-encounter removed, one with an $l>2$ encounter reduced by 1 and one where the $l$ encounter breaks into two separate ones. Reversing the steps, for any $d'$-quadruplet we can connect the two distinct encounters before the two end points by pulling out an 2-encounter (which we then remove) to give a contribution of $-D'_{K-1}$ to $\tilde{X}_K$. The minus sign derives from the final diagram having one encounter less.
Then, for any $d$-quadruplet where $o_2$ is connected to an encounter, we can join the link to $o_1$ into the encounter giving a contribution of $D_{K-1}-C_{K-1}$. Here we remove the cases where $o_2$ connects directly to $i_2$. Swapping the roles of $o_2$ and $o_1$ gives a factor of 2. Finally we may connect the final links of any $d$-quadruplet with a 2-encounter creating an $\tilde{x}$-quadruplet with an additional encounter (and 2 extra links) and hence a contribution of $-D_{K-1}$ to $\tilde{X}_K$.
Putting it all together,
\begin{eqnarray} \label{XDrel}
\tilde{X}_K &=& - D_{K-1} + 2D_{K-1}-2C_{K-1} - D'_{K-1} \nonumber \\
&=& \tilde{D}_{K-1} - \delta_{K,1} \, ,
\end{eqnarray}
using \eref{Drel}.

Since the only diagram for $K=1$ is in \fref{fig:leadingorderquads}(a), we have $\tilde{X}_1=-1$ and $\tilde{D}_1=0$, so that $\tilde{X}_K=-1$ for odd $K$ while $\tilde{D}_K=-1$ for even $K$ and both are 0 otherwise. (Both are also 0 for $K=0$).

We also need to consider the cases when an encounter of a diagram can be moved into an incoming lead. This is only possible if both incoming channels are connected
to the same 2-encounter. If, for example, an $\tilde{x}$-quadruplet starts with such a 2-encounter, then this encounter can be cut out (by moving the incoming channels to after the encounter) leaving a $\tilde{d}$-quadruplet with a value of $L-V$ which is one smaller. [The only exception is \fref{fig:leadingorderquads}(a) where the removal of the
2-encounter leads to the $d$-quadruplet in \fref{fig:leadingorderquads}(b)].
Reversing the process, all such $\tilde{x}$-quadruplets can be built by adding a 2-encounter to the front of any $\tilde{d}$-quadruplet [or \fref{fig:leadingorderquads}(b)]. Similarly, one can interchange the roles of the $\tilde{x}$- and $\tilde{d}$-quadruplets and create any $\tilde{d}$-quadruplet starting with a 2-encounter by adding a 2-encounter to the front of any $\tilde{x}$-quadruplet with a value of $L-V$ which is smaller by one. If we denote by an undertilde quadruplets with an encounter that can be moved into a lead then we have
\begin{equation} \label{eq:undertildequads}
\utilde{\tilde{X}}_K = - \tilde{D}_{K-1} \, , \quad  \utilde{\tilde{D}}_K = - \tilde{X}_{K-1} \, , \quad K > 1 \, ,
\end{equation}
and $\utilde{\tilde{X}}_1 = - D_0$.

\subsection{The second moment of the proper time-delays} \label{sec:m2unit}

We now return to the calculation of $m_2$. It is based on the $\tilde{x}$-quadruplets. For each quadruplet we have $L+2$ links, $V$ encounters and a factor of $M$ for each channel. Both end points can also be moved into the last encounter, and one encounter can possibly be moved into an incoming lead, giving a combined contribution of
\begin{equation} \fl
\frac{M^{V+2}}{M^{L+2}} (-1)^{V-2}(-1+\delta) \, (-1+2)   = - \frac{(-1)^{V}}{M^{L-V}} (1 - \delta) \, , \quad \mbox{if} \; \; \; L-V >1 \, ,
\end{equation}
where $\delta = 1$ if an encounter can move into a lead, and $0$ otherwise. The case $L-V=1$ is different, because then we
have the diagram in \fref{fig:leadingorderquads}(a) where the same encounter can receive the end points and move into a lead. Then the brackets $(-1+\delta)(-1+2)$ are replaced by $(-2)$. When we sum over all diagrams using the results of section \ref{sec:unitarymanipdiag}, using \eref{eq:undertildequads} for the $\delta=1$ case (and further multiply by $T_H^2/M=M/\mu^2$), we have
\begin{eqnarray} \label{m2unitresult}
\mu^2 m_2 &=& - 2 \tilde{X}_1 - \sum_{K=2}^{\infty} \frac{\tilde{X}_K}{M^{K-1}} - \sum_{K=2}^{\infty} \frac{\tilde{D}_{K-1}}{M^{K-1}} \nonumber \\
& = & 2 \sum_{k=0}^{\infty} \frac{1}{M^{2k}} = \frac{2M^2}{M^2-1} \, ,
\end{eqnarray}
where the term $-2 \tilde{X}_1$ gives the leading order result in \eref{secondmomentleadingorder}. The final expression in \eref{m2unitresult} is exactly the RMT result \eref{eq:m2} at $\beta=2$.

\subsection{Variance of the Wigner time delay} \label{sec:vartdunit}

This task requires the second moment of a different type, $\langle\tW^2\rangle=\frac{1}{M^2}\langle[\Tr Q]^2\rangle$. The trajectories belonging to ${\boldsymbol \gamma'}$
connect now $i_1$ to $o_1$ and $i_2$ to $o_2$ and hence the $d$-quadruplets are the relevant diagrams.  We note that for computing the variance we only need to consider connected diagrams, since the remaining diagrams merely cancel the average time delay squared (in fact, the first $d$-quadruplet in \fref{fig:leadingorderquads}(b) does this). Compared to the calculation for the moment $m_2$ in the previous section, the role of the $\tilde{x}$- and $\tilde{d}$-quadruplets are simply reversed, the case $K=1$ does not contribute, and due to the normalisation of the variance we don't have to multiply by $M/\mu^2$.
The variance of the Wigner time delay is therefore given by the sum
\begin{eqnarray} \label{vartdunitresult} \fl
\var(\tW) = \frac{1}{\bar{\tau}^2_W} \left\langle (\tau_W - \bar{\tau}_W)^2 \right\rangle
&=& - \sum_{K=2}^{\infty} \frac{\tilde{D}_K}{M^{K}} - \sum_{K=2}^{\infty} \frac{\tilde{X}_{K-1}}{M^{K}} \nonumber \\
&=& 2\sum_{k=1}^{\infty} \frac{1}{M^{2k}} = \frac{2}{M^2-1} \, .
\end{eqnarray}
This result fully agrees with RMT, as can be seen by setting $\beta=2$ in \eref{eq:tWvar}.

\subsection{Variance of the partial time-delays}

As already discussed in the beginning of \sref{sec:approach}, for systems without TRS the statistics of the partial time-delays turns out to be equivalent (at perfect coupling) to those of the diagonal elements, $q_c$. Hence, we can consider $\var(q_c)$ which involves pairs of trajectories going to different end points but all starting in the same channel.  Since the incoming channels coincide, the $\tilde{d}$- and $\tilde{x}$-quadruplets now all have the same number of free channels and contribute at the same order.  The leading order $d$-quadruplet cancels the mean part squared [as for $\var(\tW)$], leaving
\begin{eqnarray} \label{varqunitresult}
\var(q_c) &=& - \sum_{K=1}^{\infty} \frac{\tilde{X}_K}{M^{K}} - \sum_{K=1}^{\infty} \frac{\tilde{D}_K}{M^{K}} \nonumber \\
& = & \sum_{k=1}^{\infty} \frac{1}{M^{k}} = \frac{1}{M-1} \, .
\end{eqnarray}
This is in line with the RMT result \eref{eq:tcvar} at $\beta=2$, see also \ref{app:unitpartial} for an alternative calculation of $\var(t_c)$ explicitly.

\subsection{Time-reversal symmetry} \label{sec:trs}

By creating the relations above between the $\tilde{x}$- and $\tilde{d}$-diagrams, we could derive the second moments without any recourse to cutting periodic orbit diagrams.
This is notably simpler than the results for the standard transport second moments (\eg shot noise and the conductance variance) \cite{mulleretal07}. Treating the case with TRS requires, however, a more involved calculation which we pursue in the Appendices. First, we map the problem without TRS to periodic orbits in \ref{app:unitrelation} and then evaluate the semiclassical sums which arise in \ref{app:unitsums}. The corresponding sums for systems invariant under time-reversal are worked out in \ref{app:orthosums}. This allows us to finally obtain, at any $M$, semiclassical results for the second moments of time-delays in the case of orthogonal symmetry in \ref{app:orthsecondmoments}.  We find identical results to RMT for the variance of the Wigner time delay and the second moment of the proper time-delays, and derive a new result for the variance of the diagonal elements (which in the case of orthogonal symmetry is not the same as the variance of the partial time-delays). We also demonstrate the equivalence of the semiclassical approach for the time-delay problem developed here to previous treatments in \ref{app:comparison}.

%%%%%%%%%%%%%%%%%%%%%%%%%%%%%%%%%%%%%%%%%%%%%%%%%%%%%%

\addtocontents{toc}{\vspace{-1em}}
\section{Moment generating functions} \label{sec:momgenfuncs}

\begin{figure}
  \centering
  \includegraphics{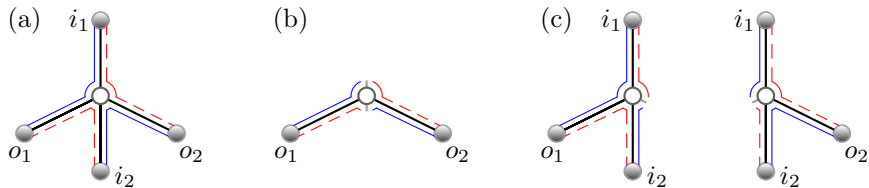}
  \caption{The diagram from \fref{fig:leadingorderquads}(a) can be redrawn as the rooted plane tree in (a).  The encounter becomes the circle in the middle and the trajectory quadruplet now travel around the outside of the tree whose leaves alternate between incoming and outgoing channels.  Since the encounter is connected to two $i$-leaves it may move into the incoming lead leaving the diagram in (b) which corresponds to \fref{fig:secondmoments}(a).  Alternatively either outgoing channel can move into the encounter as in (c) which are a new representation of the diagrams in \frefs{fig:secondmoments}(b) and (c).}
  \label{fig:treepic}
\end{figure}

We now apply the diagrammatic rules established in \sref{sec:diagrules} to derive an expansion in inverse channel number of the moment generating function of proper time-delays,
\begin{equation}
G(s) \equiv \sum_{n=1}^{\infty} \mu^n s^n m_n = G_0 +\frac{G_1}{M} + \ldots\,.
\end{equation}
As shown in \cite{bhn08}, and further developed in \cite{bk10,kuipersetal11}, the contribution of leading order in $M^{-1}$ comes from diagrams that can be represented as rooted plane trees and which can therefore be constructed recursively.  For example, the diagram in \fref{fig:leadingorderquads}(a) can be redrawn as the boundary walk around the tree in \fref{fig:treepic}(a).

Rooted trees can of course be considered as unrooted trees, which we will call subtrees below, appended to a single point.  Going beyond the leading order amounts instead to adding subtrees to increasingly intricate \emph{base structures}, following the formalism of \cite{bk11}.  Leaving the details of the diagrammatic approach to \cite{bk11}, we highlight below how it can be adapted to the  semiclassical method presented here.

\subsection{Subtrees}

Starting with unrooted trees, there are two types: one with an excess of outgoing leaves whose generating function we call $f$ and one with a excess of incoming leaves counted in $\fh$.  For example, removing the top $i_1$ channel from the tree in \fref{fig:treepic}(a) to obtain an unrooted tree, an excess of $o$-leaves remains forming a subtree included in $f$.
The generating variable will be $r$ whose power counts the number of leaves, which is related to the order of the moment.
Breaking the trees at the top encounter node gives the recursion
\begin{equation}
f = r - \sum_{l=2}^{\infty}f^l \fh^{l-1} + r\sum_{l=2}^{\infty}l f^{l-1} \fh^{l-1} \, ,
\end{equation}
where the first term is an empty tree going straight to an outgoing leaf, the next term are trees with encounter nodes of size $l$ with $l\geq2$ and the last term is the correction for allowing one outgoing leaf to move into the encounter (and replacing the corresponding general tree $f$) where we have the factor of $r$ to account for the lost leaf.
Summing we get
\begin{equation}
\frac{f}{(1-h)} = r \frac{\partial}{\partial h} \frac{1}{(1-h)} = \frac{r}{(1-h)^2} \, ,
\end{equation}
with $h=f\fh$.
This also gives us the useful relation
\begin{equation} \label{rsimpeq}
r=f(1-h)  \, .
\end{equation}

The trees of type $\fh$ with an excess of incoming leaves can actually also move into the incoming lead if all the $\fh$ type trees after the top encounter end directly in an incoming channel.
Then we have the recursion
\begin{equation}
\fh = r - \sum_{l=2}^{\infty}\fh^l f^{l-1} +r\sum_{l=2}^{\infty}(l-1) \fh^{l} f^{l-2} + \sum_{l=2}^{\infty}r^l f^{l-1} \, ,
\end{equation}
where the first three terms again correspond to an empty tree, trees with top encounter of size $l$ and moving outgoing leaves into the encounter while the last term is the new possibility of moving the encounter into the incoming lead.
Summing gives
\begin{equation} \label{phsumeq}
\frac{\fh}{(1-h)} = \frac{r\fh^2}{(1-h)^2} + \frac{r}{(1-rf)} \, .
\end{equation}
Substituting for $r$ from \eref{rsimpeq} in just the first term on the right gives the simplification
\begin{equation} \label{phsumsimpeq}
\frac{\fh}{(1-h)} = \frac{h\fh}{(1-h)} + \frac{r}{(1-rf)}, \qquad \fh = \frac{r}{(1-rf)} \, ,
\end{equation}
from which we can get a quadratic for $f$, $\fh$ and more importantly $h$
\begin{equation} \label{hendeq}
h^2+(s-1)h+s=0 \, ,
\end{equation}
with $s=r^2$.

\subsection{Leading order}

To get the full leading order moments with generating function $G_0$, we need to root our trees by adding an incoming channel to the $f$ trees (and divide by $M$) or adding an outgoing leaf to the $\fh$ trees at the top.
The first option then allows the trees to also move into the incoming lead to give
\begin{equation}
G_0 = rf + \sum_{l=2}^{\infty}r^l f^{l} = \frac{rf}{1-rf}  \, .
\end{equation}
Using \eref{phsumsimpeq} this is just
\begin{equation}
G_0 = h \, .
\end{equation}

Alternatively, and more simply, we can place an outgoing leaf on top of the $\fh$ type trees
\begin{equation}
G_0 = r\fh + r\sum_{l=2}^{\infty}\fh^l f^{l-1} = \frac{r \fh}{1-h} = h \, ,
\end{equation}
where the sum is over the additional possibility of placing the new outgoing leaf into the encounter.

Either way, the end result is that
\begin{equation} \label{G0result}
G_0 = \frac{1-s-\sqrt{1-6s+s^2}}{2}\, ,
\end{equation}
from the correct solution of \eref{hendeq}.
This is exactly what was in \cite{bk10} and equivalent to the previous RMT result \cite{bfb99}.
Compared to the energy dependent cubic equations of \cite{bk10} with corresponding energy derivatives and identity matrix corrections, the result \eref{G0result} can now be obtained much more simply and directly with our new semiclassical approach for the proper time-delays.

\subsection{Subleading order} \label{sec:subleadingmomgen}

Moving to the subleading order, we can continue looking for the dominant diagrams. The non-vanishing contribution exists only for systems with TRS, whereas for those without TRS the first correction occurs in the second subleading order (see below). As shown in \cite{bk11}, the relevant diagram in the former case has the topology of a M\"obius strip that arises after merging the quadruplets with time-reversal partners. Around the M\"obius strip there are two types of nodes, those with an even number of subtrees on each side and those with an odd number.  We shall denote these as an even node and odd node respectively.  For an $l$ encounter there are $(l-1)$ subtrees of each type (2 stretches are the M\"obius loop itself) which can be arranged in $l$ ways with an even number on each side and $(l-1)$ ways with an odd number on each side. These nodes cannot move into the incoming lead, but directly connecting odd leaves can be moved into the encounter node (one at a time). Using slightly different notation than \cite{bk11} which is more useful for the higher orders in \sref{sec:algoresults} and \ref{app:algorithmic}, the contribution of an even node is
\begin{equation}
\fl \mathcal{A} = -\sum_{l=2}^{\infty} l h^{l-1} + r \fh \sum_{l=2}^{\infty} l(l-1) h^{l-2} = \frac{h(h-2)}{(1-h)^2}+\frac{2r\fh}{(1-h)^3} = \frac{h^2}{(1-h)^2} \, ,
\end{equation}
while an odd node contributes
\begin{equation}
\fl \mathcal{B} = -\sum_{l=2}^{\infty} (l-1) h^{l-1} + r \fh \sum_{l=2}^{\infty} (l-1)^2 h^{l-2} = -\frac{h}{(1-h)^2}+\frac{r\fh(1+h)}{(1-h)^3} = \frac{h^2}{(1-h)^2} \, .
\end{equation}
Both include the link leading up to the encounter node (but not the one leaving as this is included in the next node).

Around the M\"obius loop we have an arbitrary number of encounter nodes, but we need to divide by their number because of the rotational symmetry.  As such we define a generating function of an arbitrary arrangement of nodes around a loop
\begin{equation}
\tilde{\mathcal{K}}_1 = \frac{1}{2}\sum_k \frac{(\mathcal{A}+p\mathcal{B})^k}{k}
   = -\frac{1}{2}\log \left(1-\mathcal{A}-p\mathcal{B}\right)  \, ,
\end{equation}
where we divide by 2 to account for the inside/outside symmetry of the loop.  A factor $p$ is included with the odd nodes since we actually need to have an odd number of odd nodes to close the loop properly.  This is then achieved by comparing the values of $\tilde{\mathcal{K}}_1$ at $p=\pm1$, giving
\begin{equation}
 \mathcal{K}_1 = \frac{\tilde{\mathcal{K}}_1(p=1)-\tilde{\mathcal{K}}_1(p=-1)}{2}
 = -\frac{1}{4}\log \left(1-\frac{2h^2}{(1-h)^2}\right)  \, ,
\end{equation}
which is the integrated moment generating function.
For the generating function itself, we differentiate
\begin{equation}
G_1^{\OO} = 2s \frac{\rmd\mathcal{K}_1}{\rmd s} \, ,
\end{equation}
so that by using the solution for $h$ from \eref{hendeq} we get the result
\begin{equation}
G_1^{\OO} =  \frac{1-3s-\sqrt{1-6s+s^2}}{2(1-6s+s^2)} \, ,
\end{equation}
which is the same as in \cite{bk11} but obtained in much more straightforward way, without needing to add and then remove an energy dependence.

\subsection{Algorithmic approach} \label{sec:algoresults}

To treat higher-order corrections, we can use the algorithmic approach of \cite{bk13b} which makes use of combinatorial base structures built on permutations describing the diagram's edges and vertices.
The possible permutations are generated via a computer search, while the permissible edge and vertex components are matched up according to the prescription of the permutation.  All that is then needed by the algorithm are the general semiclassical contribution of the possible edge and vertex components which we list in \ref{app:algorithmic}.

Here instead we merely state the results, which for unitary symmetry are
\begin{equation}
G_2^{\U} = \frac{2s^2}{(1-6s+s^2)^{\frac{5}{2}}} \, ,
\end{equation}
confirming the guess in \cite{bk11}, and
\begin{equation}
G_4^{\U} = \frac{2s^2(1+30s+3s^2-12s^3+8s^4)}{(1-6s+s^2)^{\frac{11}{2}}} \, .
\end{equation}

For the orthogonal case (with TRS), the results are
\begin{equation}
\fl G^{\OO}_{2}(s) = \frac{s(s-3)}{\left(1-6s+s^2\right)^2}+\frac{3s(s-1)^2+2s^2}{\left(1-6s+s^2\right)^{\frac{5}{2}}} \, ,
\end{equation}
again confirming the guess in \cite{bk11}, while at higher order we have
\begin{equation}
\fl G^{\OO}_{3}(s) = -\frac {2s\left(6{s}^{4}-5{s}^{3}+9{s}^{2}-15s-3\right)}
{\left(1-6s+{s}^{2}\right)^{4}}
-\frac {2s\left(3+19s-9{s}^{2}+2{s}^{3}\right)}
{\left(1-6s+{s}^{2}\right)^{\frac{7}{2}}} \, ,
\end{equation}
and
\begin{eqnarray}
\fl G^{\OO}_{4}(s) & = & \frac {4s\left(6{s}^{4}-30{s}^{4}+123{s}^{3}-147{s}^{2}-85s-3\right)}
{\left(1-6s+{s}^{2}\right)^{5}} \nonumber \\
\fl & & +\frac{2s\left(6+163s+216{s}^{2}-219{s}^{3}+24{s}^{4}+20{s}^{5}+36{s}^{6}\right)}
{(1-6s+s^2)^{\frac{11}{2}}} \, .
\end{eqnarray}

At the highest two orders for both symmetry classes, the generating functions are new results not yet calculated semiclassically \cite{bk11} or using RMT \cite{mezz12}.

%%%%%%%%%%%%%%%%%%%%%%%%%%%%%%%%%%%%%%%%%%%%%%%%%%%%%%%%%%%%%

\addtocontents{toc}{\vspace{-1em}}
\section{Conclusions and discussion}\label{sec:concs}

An efficient method is developed for the semiclassical calculation of the statistics of time delays in chaotic cavities. The method relies on the resonant representation of the Wigner-Smith time-delay matrix that has the advantage of not involving an energy derivative. It can be expressed in terms of semiclassical trajectories that enter the system and terminate inside. Under spectral averaging, the results for time-delay moments are then produced by sums over sets of classical trajectories which can be evaluated using simple diagrammatic rules. For \emph{individual systems} we establish in this way the universality of the RMT predictions for the second moments, including the variance of the Wigner time delay, at arbitrary number of open channels.

We also significantly advance the computation of the moment generating function of the proper time-delays, for which the first five orders are found in \sref{sec:algoresults}. This has been achieved by incorporating the present approach into the algorithmic formalism of \cite{bk11}, leading to much simpler diagrammatic rules for the trees which arise.  Although previous semiclassical approaches involving  energy correlations can be used to obtain the same results, the quadratic subtree equations become cubic, making the derivations and semiclassical contributions notably more involved. For the second moment of the proper time-delays and the variance of the Wigner time delay, there are tricks to reduce the difficulty of the previous  semiclassical treatment, as discussed in \ref{app:comparison}.  However, the results can only be obtained using the sums evaluated here in \ref{app:unitsums} and \ref{app:orthosums} while the variance of the diagonal elements of the time-delay matrix cannot be treated.  The new approach developed here presents a simpler and, more importantly, a unified approach to all the second moments. Notably for the unitary case, the second time-delay moments can be calculated without any recursive sums, which is not even possible for their transport analogues.

But the main advantage of our approach is that the diagrams and their contributions are now more similar to those of quantum transport problems, as developed in \cite{mulleretal05}. Transport moments are expressed in terms of the transmission eigenvalues that follow the Jacobi ensemble of RMT \cite{beenakker97,bb96}. Note that arbitrary transport moments can also be expressed in terms of recursively generated functions called Weingarten coefficients.  Moreover, the corresponding semiclassical diagrams can be mapped to certain types of primitive factorisations \cite{bk12, bk13a} which in turn match the Weingarten coefficients so that semiclassics and the Jacobi ensemble can be proven to be identical \cite{bk13a}.  The time-delay matrix (in its `symmetrised' form \cite{bfb97}) follows, however, an inverse Wishart distribution and its eigenvalues form the Laguerre ensemble \cite{bfb99}.  Intriguingly, the moments of inverse real Wishart matrices, corresponding to time delays with TRS, were recently expressed in terms of deformed Weingarten coefficients \cite{matsumoto12}.  Indeed, appropriately substituting $\gamma=M/2$ and $\sigma^{-1} = \tH I/2$ into the example formulae in \cite{matsumoto12}, one quickly obtains the second moments in \eref{eq:tWvar}, \eref{eq:tcvar} and \eref{eq:m2} with $\beta=1$.  Therefore, the new semiclassical approach, with its simplified diagrammatic rules, opens the door for a dynamical justification of the use of RMT and the Laguerre ensemble in the time-delay problem.

To this end, we mention that another method was recently put forward by Novaes \cite{nova13c} who suggested to use a matrix model which generates the same diagrams as in semiclassics but can be calculated exactly in RMT. For systems with broken TRS, the method was originally implemented to the transport problem, yielding successfully the exact RMT results for general counting statistics, and then further applied to time delays \cite{nova14}, producing the exact moments of proper time-delays up to 8th order semiclassically. However, showing the full equivalence for arbitrary moments was not yet possible, mainly due to the diagrammatic complexity induced by energy correlations. We believe that the incorporation of the semiclassical approach developed here into the formalism of \cite{nova14} is a promising way to go forward in establishing the full RMT-semiclassics equivalence. Further developments include generalizations to the case of preserved TRS as well as other symmetry classes, \eg Andreev billiards for which the statistics of the time-delay matrix was recently derived in \cite{marc14}.

Throughout this article, we have considered the universal regime described by RMT, which neglects the effects due to a finite Ehrenfest time \cite{alei96}. The latter is responsible for system-specific corrections which can only be obtained semiclassically \cite{ada03}; various applications to transport were already discussed in \cite{jw06,br06a,wj06} and more recently in \cite{wkr11,wkjr12}. The semiclassical representation developed here is already tailored for taking into account such effects in the time-delay problem, calling for further study in this direction. We also mention the challenge of generalising the approach to treat  other `real-world' effects, \eg absorption \cite{ss03} and non-ideal coupling \cite{kr13}.
\\[1ex]
\textbf{Acknowledgement}
\\[1ex]
Support from the funding programme Open Access Publishing from the German Research Foundation (DFG) is gratefully acknowledged.

%%%%%%%%%%%%%%%%%%%%%%%%%%%%%%%%%%%%%%%%%%%%%%%%

\appendix %
\renewcommand\thesubsection{\Alph{section}.\arabic{subsection}}

%\addtocontents{toc}{\vspace{-1em}}
\section[\hspace{5em}Partial time-delays without time-reversal symmetry]{Partial time-delays without time-reversal symmetry} \label{app:unitpartial}

A partial time-delay may be related to the matrix $Q$ as follows \cite{savi01}
\begin{equation} \label{tcdef}
  t_c = [U^{\dag}QU]_{cc} = \sum_{a,b} Q_{ab} U_{bc}U_{ac}^{*}\,,
\end{equation}
where $U$ is the matrix which diagonalises the scattering matrix $S$.  First, we note that the averages over $Q$ and $U$ can be taken independently for the unitary case \cite{bfb97}.  Then, averages over $U$ from the CUE are known both from RMT \cite{bb96,mell90} and semiclassically \cite{bk13a}.  Combined together, this readily yields the mean time delay
\begin{equation} \label{tcunitaverage}
\fl \left\langle t_c \right\rangle  = \frac{1}{M}\sum_{a,b} \left \langle Q_{ab} \right\rangle \delta_{a,b} = \frac{1}{M}\left \langle \sum_{a} Q_{aa} \right\rangle = \frac{\left\langle \Tr Q \right \rangle}{M} = \tW \, .
\end{equation}
For a partial time-delay squared,
\begin{equation} \label{tcsquare}
 t_c^{2}= \sum_{\substack{a_1,b_1\cr a_2,b_2}} Q_{a_1b_1}Q_{a_2b_2} U_{b_1c}U_{b_2c}U_{a_1c}^{*}U_{a_2c}^{*} \, ,
\end{equation}
we use the known average
\begin{equation} \label{U2correlatorunit}
\left\langle U_{b_1c}U_{b_2,c}U_{a_1c}^{*}U_{a_2c}^{*} \right\rangle =\frac{\delta_{a_1,b_1}\delta_{a_2,b_2}+\delta_{a_1,b_2}\delta_{a_2,b_1}}{M(M+1)}
\end{equation}
to obtain
\begin{equation} \label{tcunitsquaredaverage}
\fl \left\langle t_c^{2} \right\rangle = \frac{1}{M(M+1)}\left\langle \sum_{a,b} Q_{aa}Q_{bb} + Q_{ab}Q_{ba} \right\rangle  = \frac{\left\langle [\Tr Q]^2 \right \rangle + \left\langle \Tr [Q^2] \right \rangle}{M(M+1)} \, .
\end{equation}
Substituting here the expressions from \eref{m2unitresult} and \eref{vartdunitresult}, we find
\begin{equation} \label{tcunitsquaredresult}
\left\langle t_c^{2} \right\rangle = \frac{\tW^2}{M(M+1)}\left(\frac{2M^2}{M^2-1} + M^2 + \frac{2M^3}{M^2-1} \right) = \frac{M\tW^2}{M-1} \, ,
\end{equation}
which gives a (rescaled) variance of $\var(t_c) = 1/(M-1)$, \ie identical to \eref{varqunitresult}.

%%%%%%%%%%%%%%%%%%%%%%%%%%%%%%%%%%%%%%%%%%%%%%%%%%%%%%%%%%

\addtocontents{toc}{\vspace{-1em}}
\section[\hspace{5em}Mapping to periodic orbit structures for the unitary case]{Mapping to periodic orbit structures for the unitary case} \label{app:unitrelation}

For the calculation of the second moments without TRS in \sref{sec:secondmoms} we did not need to resort to using periodic orbit structures.
Since the numbers $N(v)$ of periodic orbit pairs are known however we may explore the relation between orbits and quadruplets with both outgoing leaves attached to the same encounter.

\subsection{$d$-quadruplets}

To obtain an arbitrary $d$-quadruplet, one can cut any pair of links of any correlated orbit pair (including the same link twice) \cite{mulleretal07}.
If we cut links that leave different encounters however, then the resulting quadruplet will have one $o$-leaf on each encounter and the diagram's contribution will be 0.

To remain attached to the same $l$-encounter, we simply need to cut different links leaving that encounter (we also cannot cut the same link twice) which can de done in $l(l-1)$ ordered ways. The total number of $\tilde{d}$-quadruplets for each vector $v$ is then
\begin{equation}
N_{\tilde{d}}(v) = \sum_l \frac{l(l-1)v_l}{L}N(v) \, ,
\end{equation}
and so
\begin{equation}
\tilde{D}_K = \sum_v^{L-V=K} (-1)^V \sum_l \frac{l(l-1)v_l}{L}N(v) \, .
\end{equation}

\subsection{$x$-quadruplets}

To obtain an arbitrary $x$-quadruplet, one can cut out any 2-encounter \cite{mulleretal07}.
But again we only need to consider the case when both outgoing leaves are connected to the same encounter as all other cases cancel.

\begin{figure}
  \centering
  \includegraphics{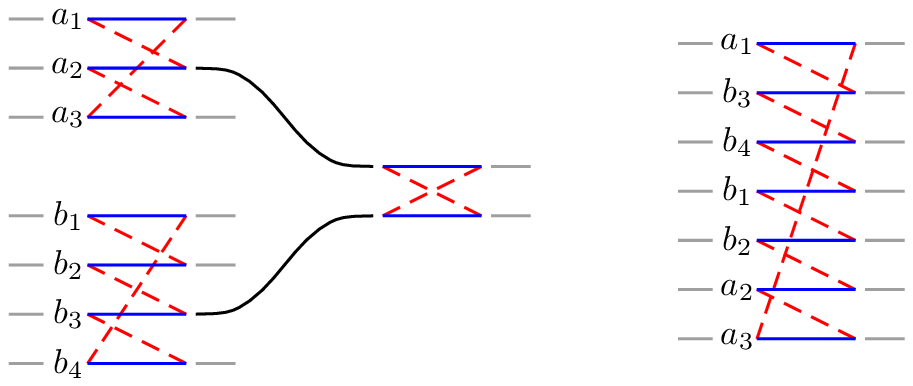}
  \caption{A periodic orbit pair with a 2-encounter connected to two different encounters.  Shrinking the links between them and the 2-encounter leads to a single larger encounter.}
  \label{fig:xcutting}
\end{figure}

Let us first consider instead the opposite case where the outgoing leaves are connected to different encounters and we have an $x'$-quadruplet.
The parts of the periodic orbit must be arranged as in \fref{fig:xcutting}.
Say that an $l_1$-encounter and an $l_2$-encounter are connected to the 2-encounter with encounter links numbered by $a_i$ and $b_i$ respectively.
When the links between the encounters and the 2-encounter are shrunk, a single $l$-encounter with $l=l_1+l_2$ is created.
If the reconnection of the original $l_1$-encounter is represented by the permutation $(a_1 \ldots a_{l_1})$ and of the $l_2$ encounter by $(b_1 \ldots b_{l_2})$ and the 2-encounter swaps links $(a_i b_j)$ then the $l$-encounter has the permutation
\begin{equation}
\fl (a_i b_j)(a_1 \ldots a_{l_1})(b_1 \ldots b_{l_2}) = (a_1 \ldots a_{i-1},b_{j}\ldots b_{l_2} b_1 \ldots b_{j-1}, a_i \ldots a_{l_1})  \, .
\end{equation}

Reversing the process, we may pull a 2-encounter out of an $l$-encounter (and break it into an $l_1$- and $l_2$-encounter) if we can turn the cycle $(1\ldots l)$ into a 2-cycle followed by a $l_1$ and $l_2$ cycle with $l_1,l_2\geq2$.
For this we can easily check that multiplying $(1\ldots l)$ on the left by $(i j)$ leads to cycles of the correct size as long as $i$ and $j$ are not equal or adjacent (cyclically).
There are then $l(l-3)$ ways of pulling a 2-encounter out of an $l$-encounter for $l>3$.
Note that this process adds two links and two encounters so the value of $L-V$ remains constant.

For any periodic orbit pair described by a vector $v$, an $x'$-quadruplet can therefore be created in $\sum_{l>3} l(l-3)v_l$ ways while $x$-quadruplets could be created in the $2v_2$ ways of cutting out any 2-encounter. Taking the difference to obtain $\tilde{x}$-quadruplets leaves
\begin{equation}
2v_2 - \sum_{l>3} l(l-3)v_l = -\sum_{l\ge2}l(l-3)v_l
\end{equation}
possibilities.

Removing a 2-encounter from a periodic orbit leaves a x-quadruplet with $L$ links, $V-1$ encounters and a vector with a 2-encounter removed. Recalling the minus contribution of encounters leads to
\begin{equation}
\tilde{X}_K = \sum_v^{L-V=K+1} (-1)^V \sum_l \frac{l(l-3)v_l}{L}N(v) \,.
\end{equation}
%

%%%%%%%%%%%%%%%%%%%%%%%%%%%%%%%%%%%%%%%%%%%%%%%%%%%%

\addtocontents{toc}{\vspace{-1em}}
\section[\hspace{5em}Evaluating unitary sums]{Evaluating unitary sums} \label{app:unitsums}

Now we turn to evaluating such sums. Since without TRS,
\begin{equation} \label{unitsumcancel}
C_K = \sum_v^{L-V=K}(-1)^V N(v) = \sum_v^{L-V=K}(-1)^V\frac{\sum_l lv_l}{L} N(v)  = 0 \, ,
\end{equation}
(for the general case of $K>0$) the sums for both the $\tilde{d}$- and $\tilde{x}$-quadruplets reduce to evaluating
\begin{equation}
H_K = \sum_v^{L-V=K}(-1)^V\frac{\sum_l l^2v_l}{L} N(v) \, .
\end{equation}
In fact from the results in \sref{sec:unitarymanipdiag}, $H_K$ must be -1 for even $K$ and 0 otherwise (for $K>0$).
Here though instead we wish to evaluate the sum directly since this will be useful for systems with TRS.

In terms of $N(v,l)=lv_lN(v)/L$ we wish to know
\begin{equation} \label{sumwewant}
H_K = \sum_v^{L-V=K}(-1)^V \sum_{l\ge2} l N(v,l) \, ,
\end{equation}
while these numbers satisfy the recursion relation \cite{mulleretal05}
\begin{eqnarray} \label{unitrecurs}
N(v,l) &=& \sum_{k\ge2}N(v^{[k,l\to k+l-1]},k+l-1) \\
& & {} + \sum_{k=1}^{l-2}(l-k-1)(v_{l-k-1}+1)N(v^{[l\to m,l-m-1]},k) \, , \nonumber
\end{eqnarray}
where $v^{[a_1,\ldots,a_m \rightarrow b_1, \ldots , b_n]}$ denotes a vector that is obtained from $v$ by decreasing all components $v_{a_i}$ by one, and increasing all $v_{b_j}$ by one. The recursions are derived by shrinking each link of the periodic orbits. Either a $k$- and $l$-encounter merge to form a $(k+l-1)$-encounter, giving the first term, or an $l$-encounter splits into a $k$- and $(l-k-1)$-encounter, giving the second term. Applying \eref{unitrecurs} to $l=2$, one finds
\begin{eqnarray}
\fl \sum_v^{L-V=K}(-1)^V 2N(v,2) &=& 2\sum_{v}^{L-V=K}(-1)^V\sum_{k\ge2}N(v^{[k,2\to k+1]},k+1) \\
\fl &=& -2\sum_{v'}^{L'-V'=K}(-1)^{V'}\sum_{k\ge2}N(v',k+1) \nonumber \\
\fl &=&  -2 \sum_{v}^{L-V=K}(-1)^{V}\sum_{k\ge3}N(v,k) \, , \nonumber
\end{eqnarray}
by relabelling the sums. Substituting this into \eref{sumwewant} yields
\begin{equation}
H_K= \sum_v^{L-V=K}(-1)^V \sum_{l\ge3} (l-2) N(v,l) \, .
\end{equation}
Since $\sum_l N(v,l)= N(v)$ this is just an example of using \eref{unitsumcancel}.

Substituting next the result for $l=3$ leads to
\begin{eqnarray}
H_K &=& \sum_v^{L-V=K}(-1)^V \sum_{l\ge4} (l-3) N(v,l) \\
& & {} + \sum_v^{L-V=K}(-1)^V N(v^{[3\to1,1]},1) \, . \nonumber
\end{eqnarray}
Substituting the result for increasing values of $l$ increases the lower limit of the first sum.
Since $N(v,l)=0$ for $l>K+1$, finally we have
\begin{equation} \label{sumsimplified}
\fl H_K= \sum_v^{L-V=K}(-1)^V \sum_{l\ge3} \sum_{k=1}^{l-2}(l-k-1)(v_{l-k-1}+1)N(v^{[l\to k,l-k-1]},k) \, .
\end{equation}

Now we consider particular terms in this sum.
The case where $l=3$ (and hence $k=1$) can be simplified to
\begin{equation}
\sum_v^{L-V=K}(-1)^V N(v^{[3\to1,1]},1) = - \sum_{v}^{L-V=K-2} (-1)^V(L+1)N(v) \, ,
\end{equation}
in terms of vectors with a lower value of $L-V$.
See \cite{mulleretal05} for details of the 1-cycles.
Likewise the cases where $l>3$ and $k=1$ or $k=l-2$ (whose results must be identical) boil down to
\begin{equation}
2 \sum_{v}^{L-V=K-2} (-1)^V \sum_l lv_l N(v) = 2 \sum_{v}^{L-V=K-2} (-1)^V L N(v) \, .
\end{equation}
Finally the more interesting case of when $k$ and $l-k-1$ are both at least 2.
This gives
\begin{equation}
- \sum_{v}^{L-V=K-2} (-1)^V \frac{\sum_j\sum_k j(v_j-\delta_{j,k})kv_k}{L} N(v) \, ,
\end{equation}
where we set $(l-k-1)=j$.
The Kronecker delta arises since the $(v_{l-k-1}+1)$ in the original equation \eref{sumsimplified} is only the number of $(l-k-1)$-encounters in $v^{[l\to k,l-k-1]}$ when $l-k-1\neq k$.
If they are equal it is 1 fewer.
Performing the sums over $j$ and $k$, this result simplifies to
\begin{eqnarray}  \label{lis2sumunit}
&& - \sum_{v}^{L-V=K-2} (-1)^V L N(v) + \sum_{v}^{L-V=K-2} (-1)^V \frac{\sum_l l^2v_l}{L} N(v) \\
& = & - \sum_{v}^{L-V=K-2} (-1)^V L N(v) +H_{K-2} \, . \nonumber
\end{eqnarray}

Combining the results from the three cases,
\begin{equation}
H_K = - \sum_{v}^{L-V=K-2} (-1)^V N(v) + H_{K-2} = -C_{K-2} + H_{K-2} \, .
\end{equation}
Since $C_{K-2}=0$ for $(K-2)>0$ then this is simply $H_K=H_{K-2}$ for $K>2$.
Furthermore, $H_2$ can be easily checked to be -1 while $H_1$ is 0, so that $H_K$ is always -1 for even $K$.
For $K=0$ with $C_0=1$ we could set $H_0=0$ in line with a sum over a 0 vector.

\subsection{Orbit interpretation}

The terms in the recursion relations above have a simple interpretation in terms of orbits.
If we take an arbitrary orbit with $L-V=K-2$, we can take any point in any of the $L$ links, combine it with any point in the $(L-1)$ remaining links or any point either side of the original point and join them together to make a new 3-encounter.
This adds 3 links and one encounter.
Including the (-1) for each encounter gives
\begin{equation}
 - \sum_{v}^{L-V=K-2} (-1)^V\frac{(L+1)L}{L}N(v) \, .
\end{equation}

Alternatively one may take a point from any of the $L$ links and move it to (before or after) any of the $L$ encounter stretches to increase the size of that encounter by 2.
This adds two more links and no new encounters
\begin{equation}
2\sum_{v}^{L-V=K-2} (-1)^V\frac{L^2}{L}N(v) \, .
\end{equation}

Finally one can pick stretches from different encounters, a $k$- and $l$-encounter say and combine them into a $k+l-1$ encounter.
This adds one link and reduces the number of encounters by 1.
To count them we can pick any two (ordered) encounter stretches in $L(L-1)$ ways and remove the $\sum_l l(l-1)v_l$ which come from the same encounter
\begin{equation}
\fl -\sum_{v}^{L-V=K-2} (-1)^V\frac{L(L-1)}{L}N(v) + \sum_{v}^{L-V=K-2} (-1)^V\frac{\sum_l l(l-1)v_l}{L}N(v) \, .
\end{equation}

\subsection{A further sum}

Later we will also need the following sum
\begin{equation} \label{Jsum}
\fl J_K= \sum_v^{L-V=K}(-1)^V \sum_l l^2v_l N(v) = \sum_v^{L-V=K}(-1)^V L \sum_{l} l N(v,l) \, ,
\end{equation}
which we now treat in the same way as $H_K$ for systems without TRS.
First for $l=2$ we have
\begin{eqnarray}
\fl \sum_v^{L-V=K}(-1)^V 2LN(v,2) &=& 2\sum_{v}^{L-V=K}(-1)^V L \sum_{k\ge2}N(v^{[k,2\to k+1]},k+1) \\
\fl &=& -2\sum_{v'}^{L'-V'=K}(-1)^{V'}(L'+1)\sum_{k\ge2}N(v',k+1) \, ,
\end{eqnarray}
following the steps in \eref{lis2sumunit} and since merging the encounters reduces the number of links by 1.
Relabelling the sums and substituting into \eref{Jsum} leads to
\begin{equation} \label{Jsumnew}
\fl J_K= \sum_v^{L-V=K}(-1)^V \sum_{l\geq3} \left[(l-2)L - 2\right]N(v,l) \, .
\end{equation}
The term for $l=3$ is
\begin{eqnarray}
\fl \sum_v^{L-V=K}(-1)^V (L-2)N(v,3) &=& - \sum_{v'}^{L'-V'=K} (-1)^{V'}(L'-1)\sum_{k\ge2}N(v',k+2)  \\
\fl & & - \sum_{v'}^{L'-V'=K-2} (-1)^{V'}(L'+1)^2 N(v') \, , \nonumber
\end{eqnarray}
since when two encounters are merged the number of links decreases by 1 while when a 3-encounter breaks into links, 3 links are lost in the end.
Substituting into \eref{Jsum} and relabelling the sum
\begin{equation} \label{Jsumnewnew}
\fl J_K= \sum_v^{L-V=K}(-1)^V \sum_{l\geq4} \left[(l-3)L - 1\right]N(v,l) - \sum_{v}^{L-V=K-2} (-1)^{V}(L+1)^2 N(v) \, .
\end{equation}
For $l=4$ we have $(L-1)N(v,4)$ inside the first sum.
We can continue to replace terms using \eref{unitrecurs} and the same steps we used for $H_K$ as long as we keep track of how the number of links changes for different terms in the recursions.
We find that as $l$ increases we continue to have this factor of $(L-1)$ and the sum reduces to
\begin{eqnarray} \label{Jsumend}
 J_K &=& - \sum_{v}^{L-V=K-2} (-1)^{V}(L+1)^2 N(v) \\
 & & + 2\sum_{v}^{L-V=K-2} (-1)^{V}L(L+1) N(v) \nonumber \\
 & & - \sum_{v}^{L-V=K-2} (-1)^{V}L^2 N(v) + \sum_{v}^{L-V=K-2} (-1)^{V} \sum_l l^2v_l N(v) \, . \nonumber
\end{eqnarray}
This result can be read off directly from the orbit interpretation if we remember to multiply by the new number of links minus two when making a new 3-encounter and instead by the new number of links minus one in the other cases.
We can now simplify the results from the three cases since the $L^2$ factors cancel.
This leaves
\begin{equation} \label{Jsumresult}
 J_K = -C_{K-2} + J_{K-2} \, .
\end{equation}
With $C_{K-2}=0$ for $(K-2)>0$ then $J_K=J_{K-2}$ for $K>2$.
Since $J_2$ is -1 and $J_1$ is 0, then $J_K$ is always -1 for even $K$.
For $K=0$ with $C_0=1$ we could also set $J_0=0$ in line with a sum over a 0 vector.

%%%%%%%%%%%%%%%%%%%%%%%%%%%%%%%%%%%%%%%%%%%%%%%%%%%%%

\addtocontents{toc}{\vspace{-1em}}
\section[\hspace{5em}Evaluating the orthogonal sums]{Evaluating the orthogonal sums} \label{app:orthosums}

With TRS, the numbers $N(v,l)$ satisfy slightly different recursion relations \cite{mulleretal05}
\begin{eqnarray} \label{orthorec}
N(v,l) &=& \sum_{k\ge2}N(v^{[k,l\to k+l-1],k+l-1}) \nonumber \\
& & {} +2 \sum_{k=1}^{l-2}(l-k-1)(v_{l-k-1}+1)N(v^{[l\to m,l-m-1]},k) \nonumber \\
& & {} + (l-1)N(v^{[l\to l-1]},l-1) \, ,
\end{eqnarray}
with an extra factor of 2 and a new term for when a link returns to the same encounter in the opposite direction. Performing the same steps as for the unitary case, one finds
\begin{eqnarray} \label{orthoHK}
H_K &=& -2 \sum_{v}^{L-V=K-2} (-1)^V N(v) + 2H_{K-2} \\
& & {} -2 \sum_{v}^{L-V=K-1} (-1)^V N(v)  + \sum_{v}^{L-V=K-1} (-1)^V \sum_l lN(v,l) \, . \nonumber
\end{eqnarray}
The first term on the second line derives from the $l=2$ case in \eref{orthorec} with the 1-cycle treated as in \cite{mulleretal05}. Since the 2-encounter is removed entirely in the end, a minus sign appears.
Otherwise for encounters with $l>2$, the encounter is merely made one smaller providing the remaining term on the second line, which is simply $H_{K-1}$.

With TRS, we further have
\begin{equation} \label{orthoCK}
\sum_{v}^{L-V=K} (-1)^V N(v) =C_K=(-1)^K \, ,
\end{equation}
so that those terms in \eref{orthoHK} cancel.
All told we have the recursion relation
\begin{equation}
H_K = H_{K-1}+2H_{K-2} \, ,
\end{equation}
while we can explicitly check that $H_1=H_2=-2$.
Looking at the eigenvalues and vectors of the matrix form
\begin{equation}
\left(\begin{array}{c}H_{K}\\H_{K+1}\end{array}\right)=\left(\begin{array}{cc}0 & 1\\1 &2\end{array}\right)\left(\begin{array}{c}H_{K-1}\\H_{K}\end{array}\right) \, ,
\end{equation}
we then get
\begin{equation}
H_K = -\frac{2}{3}\left(2^K - (-1)^K\right) \, .
\end{equation}

Including the factor of $L$ for $J_K$ and performing the same steps we have
\begin{eqnarray} \label{orthoJK}
\fl J_K &=& -2 C_{K-2} + 2J_{K-2} \\
\fl & & {} -2 \sum_{v}^{L-V=K-1} (-1)^V (L+2)N(v)  + \sum_{v}^{L-V=K-1} (-1)^V \sum_l (L-\delta_{l,2})lN(v,l) \, . \nonumber
\end{eqnarray}
The delta function arises since the $l=3$ case in the recursions starts with a value of $(L-2)$ as opposed to the $(L-1)$ for $l>3$.
This result can be rewritten in terms of known sums
\begin{equation}
\fl J_K = -2C_{K-2} + 2J_{K-2} - 2A_{K-1} - 2C_{K-1} + J_{K-1} - 2B_{K-1} \, .
\end{equation}
To proceed further we use the relation \cite{mulleretal07}
\begin{equation}
B_K = A_{K} + A_{K-1} -C_{K} \, ,
\end{equation}
and \eref{orthoCK} to obtain
\begin{equation}
J_K = J_{K-1} + 2J_{K-2} - 4A_{K-1} - 2A_{K-2} - 2(-1)^K \, .
\end{equation}
Using also that
\begin{equation} \label{orthoAK}
A_K = (-3)^K  = -3A_{K-1} \, ,
\end{equation}
we can rearrange the result to
\begin{equation}
\fl (J_K-A_K) = (J_{K-1}-A_{K-1}) + 2(J_{K-2}-A_{K-2}) - 2 (-1)^{K} \, ,
\end{equation}
and obtain a recursion relation for $I_K=J_K-A_K$.  With the starting values of $I_1=-1$ and $I_2=-5$ we have a general result of
\begin{equation} \label{orthoIK}
I_K = -\frac{8}{9}2^K -\frac{1}{9}(-1)^K -\frac{2}{3} K (-1)^K \, ,
\end{equation}
from which we can find $J_K$ by adding $A_K$ from \eref{orthoAK}.
For $K=0$ with $A_0=C_0=1$ we could also set $J_0=0$ in line with a sum over a 0 vector and hence $I_0=-1$ in agreement with \eref{orthoIK}.

%%%%%%%%%%%%%%%%%%%%%%%%%%%%%%%%%%%%%%%%%%%%%%%%%%%%%%

\addtocontents{toc}{\vspace{-1em}}
\section[\hspace{5em}Second moments with time-reversal symmetry]{Second moments with time-reversal symmetry} \label{app:orthsecondmoments}

When we turn to systems with TRS we have the complication that the encounter stretches can be traversed in either direction.
As such we can now divide the $x$-quadruplets into three groups: $\tilde{x}$ where the outgoing leaves are connected to the same encounter and those encounter stretches travel in the same direction, $\hat{x}$ where the outgoing leaves are connected to the same encounter but those encounter stretches travel in opposite directions, and $x'$ containing the remaining diagrams with the outgoing leaves connected to different encounters.
For each vector $v$ we count the number in the new group as $N_{\hat{x}}(v)$ and define
\begin{equation} \label{Xhsum}
\hat{X}_K=\sum_v^{L-V=K}(-1)^V N_{\hat{x}}(v) \, .
\end{equation}
Since all quadruplets belong to one group
\begin{equation} \label{Xorthosum}
X_{K}= \tilde{X}_K+\hat{X}_K+{X}'_K \, .
\end{equation}

Using the same notation for the $d$-quadruplets, the new group $\hat{d}$ contains those where the outgoing leaves connect to the same encounter from encounter stretches travelling in opposite directions with number $N_{\hat{d}}(v)$.
Likewise
\begin{equation} \label{Dhsum}
\hat{D}_K=\sum_v^{L-V=K}(-1)^V N_{\hat{d}}(v) \, ,
\end{equation}
\begin{equation} \label{Dorthosum}
D_{K}= \tilde{D}_K+\hat{D}_K+{D}'_K+2C_{K} - \delta_{K,0} \, .
\end{equation}

\subsection{First relations}

As for the unitary case, we can first consider adding a 2-encounter to the end of an $\tilde{d}$-quadruplet.
Since the relevant encounter stretches are traversed in the same direction, when the links to the new 2-encounter are shrunk, the steps are identical to those in \sref{sec:unitarymanipdiag} giving the same relation
\begin{eqnarray} \label{DXrelortho}
\tilde{D}_K &=& - X_{K-1} + 2X_{K-1} - X'_{K-1} \nonumber \\
&=& \tilde{X}_{K-1} + \hat{X}_{K-1} \, ,
\end{eqnarray}
albeit with an extra term arising when we replace $X'$ using \eref{Xorthosum}.
Also nothing changes when we start with an $\tilde{x}$-quadruplet giving
\begin{eqnarray} \label{XDrelortho}
\tilde{X}_K &=& - D_{K-1} + 2D_{K-1}-2C_{K-1} - D'_{K-1} \nonumber \\
&=& \tilde{D}_{K-1} + \hat{D}_{K-1} - \delta_{K,1} \, ,
\end{eqnarray}
using \eref{Dorthosum}.

\subsection{Second relation}

\begin{figure}
  \centering
  \includegraphics{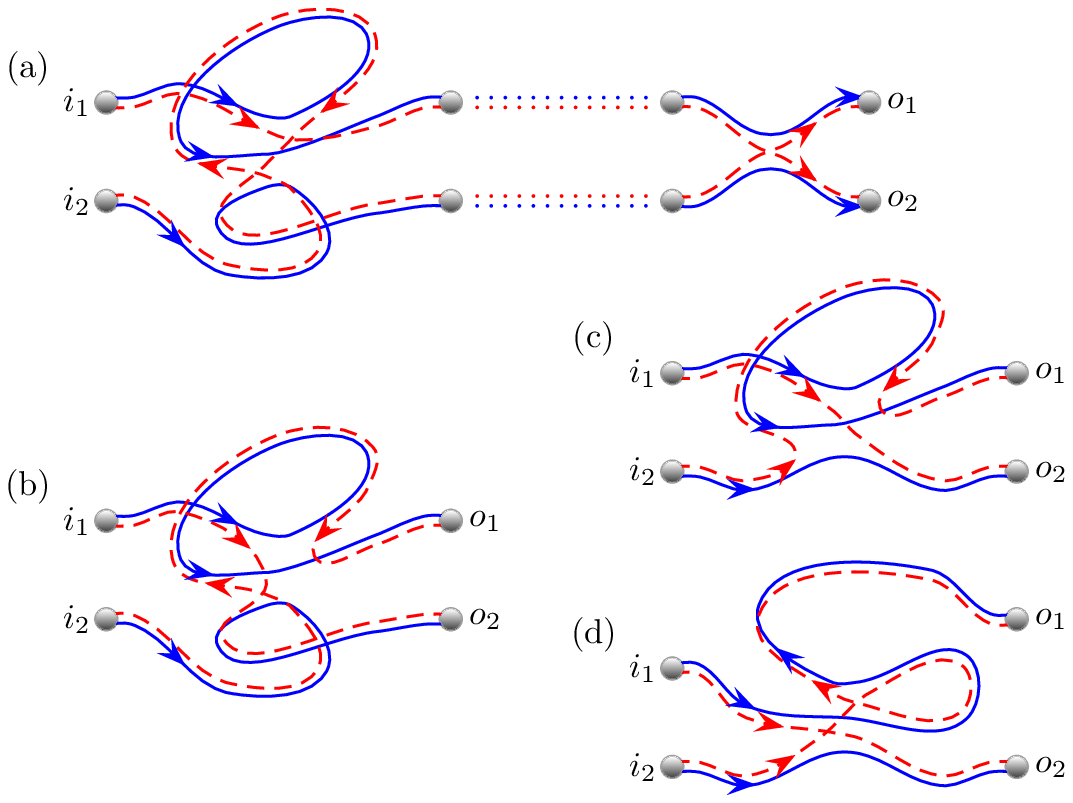}
  \caption{We start with a $\hat{d}$-quadruplet with both end links connected to the same 3-encounter, but travelling in opposite directions through the encounter.  Appending an additional 2-encounter in (a) is akin to reconnecting the dashed trajectories as in (b).  Untwisting the lower solid trajectory leads to (c) while we must also untwist the top loop so that all the encounter stretches end up parallel or anti-parallel as in (d).  This diagram is now a $\hat{x}$-quadruplet still with a 3-encounter traversed in opposite directions by the end links.  Appending another 2-encounter to (d) would effectively reverse the process and simply cancel the one added in (a).}
  \label{fig:dshrink3}
\end{figure}

Now we consider adding the same 2-encounter to the end of an $\hat{d}$-quadruplet.
Since the relevant encounter stretches are traversed in opposite directions, the $l$-encounter at the end before the new 2-encounter must have size $l>2$.
To see what happens, it is simplest to express the $l$-encounter as a permutation, and we need to record that of both the encounter stretches and their time reversals.
We use a bar to represent time reversal and hence stretches travelling in the opposite direction.
Say that stretches $a$ and $\bar{b}$ were originally connected to the outgoing leaves so that the $l$-encounter has the permutation
\begin{equation}
(ax\ldots y\bar{b}z\ldots w)(\bar{w}\ldots\bar{z}b\bar{y}\ldots\bar{x}\bar{a}) \, ,
\end{equation}
consisting of a single $l$ cycle and its time reversal.
The entries $w\ldots z$ could contain bars while the actual numbering of the elements would correspond to the order of traversal.
Adding the 2-encounter between the stretches from $a$ and $b$ corresponds to multiplying later by the permutation $(ab)$, and for the time reversal before by the permutation $(\bar{a}\bar{b})$ leaving us with
\begin{eqnarray}
(ab)(ax\ldots y\bar{b}z\ldots w)(\bar{w}\ldots\bar{z}b\bar{y}\ldots\bar{x}\bar{a})(\bar{a}\bar{b}) && \nonumber \\
= (ax\ldots y\bar{b}\bar{w}\ldots\bar{z})(z\ldots wb\bar{y}\ldots\bar{x}\bar{a}) &&  \, ,
\end{eqnarray}
again a single $l$-encounter with both outgoing leaves still attached to stretches travelling in opposite directions. Adding a 2-encounter however changes a $d$-type quadruplet into an $x$-type and once the connecting links are shrunk we still have the same number of links and encounters.  We illustrate this process in \fref{fig:dshrink3}.
Performing the same steps to a $\hat{x}$-quadruplet we just go back to a $\hat{d}$-one so this is an involution and
\begin{equation} \label{DXhatrelortho}
\hat{D}_K = \hat{X}_{K} \, .
\end{equation}

\subsection{Third relation}

So far these relations are not sufficient to determine the four quantities $\tilde{D}$, $\hat{D}$, $\tilde{X}$, $\hat{X}$ but to proceed we can use TRS to our advantage.
With TRS, we can connect the outgoing leaves $o_1$ to $o_2$ together, and also the incoming channels $i_2$ to $i_1$ and create a periodic orbit.
If we start with a $\tilde{d}$-quadruplet the link formed from connecting the outgoing leaves must return to the same encounter, but now in the opposite direction.
Starting the partner orbit in the same direction it will also follow the link made from connecting the incoming channels in the same direction.
Performing the same steps to a $\tilde{x}$-quadruplet we again have a link connecting an encounter to itself (but returning to the encounter in the opposite direction), as well as one where the orbit and its partner travel in different directions.

Given all the periodic orbits with a vector $v$ we can cut any link which connects an encounter to itself (traversed in the opposite direction) and place the outgoing leaves there.
Then we can cut any of the remaining $L-1$ links, placing the incoming channels appropriately, to create $N_{\tilde{x}}(v)+N_{\tilde{d}}(v)$.
Fortunately the number of links connecting encounters to themselves (in the opposite direction) corresponds to the third line in \eref{orthorec}.
When we multiply by the required $(L-1)$ (which cancels with the denominator) this leads to
\begin{eqnarray} \label{DXtildesumortho}
\fl \tilde{D}_K + \tilde{X}_{K} &=& \sum_{v}^{L-V=K}(-1)^V\sum_{l}(l-1)^2(v_{l-1}+1)N(v^{[l\to l-1]}) \nonumber \\
\fl &=& - \sum_{v'}^{L'-V'=K-1}(-1)^{V'}(L'+1)N(v') + \sum_{v'}^{L'-V'=K-1}(-1)^{V'}\sum_{l'}(l')^2 v_{l'}N(v') \nonumber \\
\fl &=& -A_{K-1} +J_{K-1} = I_{K-1} \, ,
\end{eqnarray}
by relabelling sums and treating the 1-cycles that arise from the $l=2$ term as in \cite{mulleretal05}.
The $l=2$ term in the first line gives the first sum on the second line which has a simple orbit interpretation.
Given an orbit with a vector $v'$ with $L'$ links, we can replace any of those links by a 2-encounter that returns to itself, adding two more links and one more encounter.
Cutting the new link returning to the 2-encounter, $L'+1$ other links can be cut to create a $\tilde{x}$- or $\tilde{d}$-quadruplet.

\subsection{Fourth relation}

Likewise, if we connect the outgoing leaves of a $\hat{x}$- or $\hat{d}$-quadruplet we arrive at a periodic orbit where a link connects an encounter to itself, but now in the same direction.
From the periodic orbits with a vector $v$ we therefore cut any link which connects an encounter to itself in the same direction and then cut any of the remaining $L-1$ links to create $N_{\hat{x}}(v)+N_{\hat{d}}(v)$.
The number of links connecting encounters to themselves (in the same direction) corresponds to the second line in \eref{orthorec}.
Again the factor of $(L-1)$ cancels and
\begin{eqnarray}
\fl \hat{D}_K + \hat{X}_{K} &=& 2\sum_{v}^{L-V=K}(-1)^V\sum_{l}\sum_{m=1}^{l-2}(l-m-1)(v_{l-m-1}+1)N(v^{[l\to m, l-m-1]}) \nonumber \\
\fl &=& -2\sum_{v'}^{L'-V'=K-2}(-1)^{V'}(L'+2)(L'+1)N(v') \nonumber \\
\fl & & + 4\sum_{v'}^{L'-V'=K-2}(-1)^{V'}(L'+1)L'N(v') \nonumber \\
\fl & & -2\sum_{v'}^{L'-V'=K-2}(-1)^{V'}(L')^{2}N(v') + 2\sum_{v'}^{L'-V'=K-2}(-1)^{V'}\sum_{l'} (l')^2v_{l'}N(v') \nonumber \\
\fl &=& -2A_{K-2}-2C_{K-2} + 2J_{K-2} = 2I_{K-2} -2C_{K-2} \, ,
\end{eqnarray}
following the same steps as in \ref{app:orthosums}.

\subsection{Final results}

Since $\hat{D}=\hat{X}$ from \eref{DXhatrelortho} we therefore have
\begin{equation} \label{DXhatorthoresult}
\hat{D}_K = \hat{X}_{K} = I_{K-2} -C_{K-2} = I_{K-2} - (-1)^{K} \, ,
\end{equation}
an explicit result for these quantities using \eref{orthoIK}.
For $\tilde{D}$ and $\tilde{X}$ we first consider the difference
\begin{equation} \label{DXtildediffortho}
\tilde{X}_{K}-\tilde{D}_K = (\tilde{D}_{K-1}+\hat{D}_{K-1}) - (\tilde{X}_{K-1}+\hat{X}_{K-1}) \, ,
\end{equation}
using \eref{DXrelortho} and \eref{XDrelortho} for $K>1$.
Again since $\hat{D}=\hat{X}$, this simplifies to
\begin{equation} \label{DXtildediffortho2}
\tilde{X}_{K}-\tilde{D}_K = -(\tilde{X}_{K-1}-\tilde{D}_{K-1}) \, .
\end{equation}
Next $(\tilde{X}_{1}-\tilde{D}_1)$ can be checked to equal $-1$ so that
\begin{equation} \label{DXtildedifforthoresult}
\tilde{X}_{K}-\tilde{D}_K = (-1)^{K} \, ,
\end{equation}
for $K>0$.  Note that for $K=0$ both $\tilde{X}_{0}$ and $\tilde{D}_0$ are 0. Substituting into \eref{DXtildesumortho} gives the explicit formulae
\begin{eqnarray} \label{DXtildeorthoresult}
2\tilde{D}_K = I_{K-1} - (-1)^{K} \nonumber \\
2\tilde{X}_K = I_{K-1} + (-1)^{K}
\end{eqnarray}
in terms of \eref{orthoIK}.
Useful for the second moments are the sums
\begin{eqnarray} \label{DXorthosums}
\tilde{D}_{K+1}+\hat{D}_{K+1}+\tilde{X}_{K}+\hat{X}_{K} &=& -\frac{4}{3}\left[2^{K}-(-1)^{K}\right] \nonumber \\
\tilde{X}_{K+1}+\hat{X}_{K+1}+\tilde{D}_{K}+\hat{D}_{K} &=& -\frac{2}{3}\left[2\cdot 2^{K}+(-1)^{K}\right] \, .
\end{eqnarray}

Finally, we can again consider the case where both incoming channels are connected to the same 2-encounter which can then be moved into the lead so that both incoming channels coincide.  As for systems without TRS in \sref{sec:unitarymanipdiag} such diagrams can be generated by simply appending a 2-encounter to the start of appropriate quadruplets.  We therefore again have the relation \eref{eq:undertildequads} while for the $\hat{x}$ and $\hat{d}$ cases we analogously have
\begin{equation} \label{eq:undertildehatquads}
\utilde{\hat{X}}_K = - \hat{D}_{K-1} \, , \quad  \utilde{\hat{D}}_K = - \hat{X}_{K-1} \, , \quad K > 1 \, .
\end{equation}

\subsection{The second moments} \label{sec:secondmomsortho}

For the calculation of $m_2$ the contribution of each diagram is as in \sref{sec:m2unit}.
When we sum over all diagrams (and further divide by $\mu^2 M$) we have to evaluate
\begin{eqnarray} \label{m2orthoresult}
\mu^2 m_2 &=& 2 - \sum_{K=2}^{\infty} \frac{\tilde{X}_K+\hat{X}_K}{M^{K-1}} - \sum_{K=1}^{\infty} \frac{\tilde{D}_K+\hat{D}_K}{M^{K}} \nonumber \\
& = & 2 - \sum_{K=1}^{\infty} \frac{\tilde{X}_{K+1}+\hat{X}_{K+1}+\tilde{D}_K+\hat{D}_K}{M^{K}} \nonumber \\
& = & \frac{2}{3} \sum_{K=0}^{\infty} \frac{2\cdot2^{K}+(-1)^{K}}{M^{K}} = \frac{2M^2}{(M+1)(M-2)} \, ,
\end{eqnarray}
where the leading order term from \eref{secondmomentleadingorder} is the same as for systems without TRS, and is included as the $K=0$ term in the final sum.  As discussed in \sref{sec:diagrules} this may be viewed as either a combination of $D_0-\tilde{X}_1-\hat{X}_1$ or simply $-2\tilde{X}_1$.  The sum over $d$-quadruplets in \eref{m2orthoresult} is again derived from $x$-quadruplets where the initial 2-encounter can be moved into the incoming lead as in \eref{eq:undertildequads} and \eref{eq:undertildehatquads}.
The result in \eref{m2orthoresult} is exactly the RMT result; see expression \eref{eq:m2} at $\beta=1$.

The variance of the time delay (dividing by the overall factor of $M^2$) is given by the related sum
\begin{eqnarray} \label{vartdorthoresult}
\var(\tW) &=& - \sum_{K=1}^{\infty} \frac{\tilde{X}_K+\hat{X}_K}{M^{K+1}} - \sum_{K=1}^{\infty} \frac{\tilde{D}_K+\hat{D}_K}{M^{K}} \nonumber \\
& = & -\frac{1}{M}\sum_{K=1}^{\infty} \frac{\tilde{D}_{K+1}+\hat{D}_{K+1}+\tilde{X}_{K}+\hat{X}_{K}}{M^{K}} \nonumber \\
& = & \frac{4}{3M} \sum_{K=1}^{\infty} \frac{2^{K}-(-1)^K}{M^{K}} = \frac{4}{(M+1)(M-2)} \, ,
\end{eqnarray}
since $\tilde{D}_1=\hat{D}_1=0$.
This result again agrees with RMT; set $\beta=1$ in \eref{eq:tWvar}.

\subsection{Spin-orbit interaction}

For the symplectic symmetry class of spin $\frac{1}{2}$ particles, the semiclassical diagrams are identical to those for the orthogonal TRS case.  Spin orbit interactions are instead included as additional spin propagators along the classical trajectories \cite{zfr05,zfr05b}.  Each channel or leaf is also split into a spin up and spin down version though observables are appropriately rescaled so that the semiclassical contribution of the diagonal pair to the average time delay remains unchanged.  At subleading order, each of the diagrams in \fref{fig:leadingoff} gain an additional factor of $-\frac{1}{2}$ \cite{wb07} but they still cancel.  Continuing to all orders \cite{gutierrezetal09} all off-diagonal contributions similarly cancel.

Treating the second moment of the proper delay times, the spin semiclassical contributions of each diagram become more complicated \cite{wb07} but effectively the leading order term remains unchanged while each higher order gains a factor of $-\frac{1}{2}$.  This can be simply incorporated by substituting $M\to-2M$ in \eref{m2orthoresult} to give
\begin{equation} \label{m2symplresult}
\mu^2 m_2 = \frac{4M^2}{(M+1)(2M-1)} \, ,
\end{equation}
which is exactly the RMT result \eref{eq:m2} at $\beta=4$.  Likewise, one finds
\begin{equation} \label{vartdsymplresult}
\var(\tW) = \frac{2}{(M+1)(2M-1)} \, ,
\end{equation}
for the variance of the Wigner time delay in full agreement with RMT, see \eref{eq:tWvar}.

\subsection{Diagonal elements of the Wigner-Smith matrix and the partial time-delays}

Returning to the orthogonal case, finally we can consider
\begin{equation} \label{varqorthoresult}
\fl \var(q_c) = - \sum_{K=1}^{\infty} \frac{\tilde{X}_K+\hat{X}_K+\tilde{D}_K+\hat{D}_K}{M^{K}} = \frac{M^2+5M+2}{(M+1)^2(M-2)} \, ,
\end{equation}
which is notably different from the RMT result for a diagonal element of the symmetrised Wigner-Smith matrix (or a partial time-delay) in \eref{eq:tcvar}. In particular, semiclassics predicts $1/M$ at leading order unlike the $2/(\beta M)$ of \eref{eq:tcvar}.  The leading order term can however also be derived from an energy dependent correlator along the lines of the calculation in \cite{ks08}.  Since none of the diagrams at leading order for $\var(q_c)$ rely on TRS and we again simply obtain the unitary result $1/M$.

Of course the difference is due to correlations between the diagonal elements of $Q$ and the symmetrisation process.  To explore this in more detail, we return to the definition of a partial time-delay in \eref{tcdef} but now perform averages over $U$ using the COE results \cite{mell80,bb96,bk13a}.  For the first moment, one finds
\begin{eqnarray} \label{tcorthaverage}
\left\langle t_c \right\rangle  &=& \frac{1}{M+1}\sum_{a} \left \langle Q_{ab} \right\rangle \left(\delta_{a,b}+\delta_{a,b,c}\right) \nonumber \\
& = & \frac{\left\langle \sum_{a} Q_{aa} \right\rangle + \left\langle Q_{cc}\right\rangle}{M+1} = \frac{\left\langle \Tr Q \right \rangle+\left\langle q_c\right\rangle}{M+1}  = \tW \, ,
\end{eqnarray}
since $M\left\langle q_c\right\rangle=\left\langle \Tr Q \right \rangle$.  Now the step of treating $Q$ and $U$ independently is no longer justified for the orthogonal case, but for the first moment it is easy to show semiclassically using our new approach.  Treating the matrix elements in \eref{tcdef} semiclassically means considering four trajectories with end points mostly determined by the channel labels.  However, the pair of trajectories coming from $Q_{ab}$ end together somewhere inside the cavity instead.  Correlations between $Q$ and $U$ would involve a quadruplet of trajectories which cannot be separated into two independent pairs [these simply give the result in \eref{tcorthaverage}].  In each semiclassical diagram there must therefore be an encounter before the link ending inside the cavity.  Moving the end point into the encounter provides a second diagram, with the opposite contribution, and all possibilities simply cancel like for the average time delay.

Treating $t_c^2$ directly would involve eight trajectories, akin to a fourth moment in standard transport, and far more complicated than the second moments considered here.  We therefore do not confirm the RMT result in \eref{eq:tcvar} but instead semiclassics provides us with the complementary result for $\var(q_c)$ in \eref{varqorthoresult} which has so far not been obtained from RMT.  To gauge the level of correlations between $Q$ and $U$ we can however compute $t_c^2$ assuming their independence, for which we need
\begin{eqnarray}\label{U2correlatororth}
\fl \left\langle U_{b_1c}U_{b_2c}U_{a_1c}^{*}U_{a_2c}^{*} \right\rangle &=&\frac{\delta_{a_1,b_1}\delta_{a_2,b_2}+\delta_{a_1,b_2}\delta_{a_2,b_1} +2\delta_{a_1,b_1,a_2,b_2,c}}{M(M+3)}  \\
\fl & & + \frac{\delta_{a_1,b_1}\delta_{a_2,b_2,c}+\delta_{a_1,b_2}\delta_{a_2,b_1,c}
+\delta_{a_2,b_2}\delta_{a_1,b_1,c}+\delta_{a_2,b_1}\delta_{a_1,b_2,c}}{(M-1)^{-1}M(M+1)(M+3)} \, , \nonumber
\end{eqnarray}
giving
\begin{eqnarray} \label{tcorthsquaredaverage}
\left\langle t_c^{2} \right\rangle &=& \frac{\left\langle \sum_{a,b} Q_{aa}Q_{bb} + Q_{ab}Q_{ba} \right\rangle + 2\left\langle Q_{cc}^2\right\rangle}{M(M+3)} \nonumber \\
& & + \frac{2(M-1)\left\langle \sum_{a} Q_{aa}Q_{cc} + Q_{ac}Q_{ca} \right\rangle}{M(M+1)(M+3)} \, .
\end{eqnarray}
Semiclassically, the result for each term $b$ in the sums in the top line is the same so that the sums in the second line incur a factor of $M^{-1}$ and we can write the full result as
\begin{equation} \label{tcorthsquaredaverageagain}
\fl \left\langle t_c^{2} \right\rangle = \frac{\left(M^2+3M-2\right)\left(\left\langle [\Tr Q]^2 \right \rangle + \left\langle \Tr [Q^2] \right \rangle\right)}{M^{2}(M+1)(M+3)} + 2\frac{\left\langle q_c^2\right\rangle}{M(M+3)} \, .
\end{equation}
Using \eref{m2orthoresult}, \eref{vartdorthoresult} and \eref{varqorthoresult} leads to
\begin{equation} \label{tcorthsquaredresult}
\frac{\left\langle t_c^{2} \right\rangle}{\bar\tW^2} = \frac{M(M^2+M+2)}{(M+1)^2(M-2)} \, ,
\end{equation}
giving a (rescaled) variance of
\begin{equation} \label{tcorthvar}
\var(t_c) = \frac{M^2+5M+2}{(M+1)^2(M-2)} \, ,
\end{equation}
which is actually equal to $\var(q_c)$, see \eref{varqorthoresult}.  The difference between \eref{tcorthvar} and \eref{eq:tcvar} are thus due to the correlations between $Q$ and $U$.

%%%%%%%%%%%%%%%%%%%%%%%%%%%%%%%%%%%%%%%%%%%%%%%%%%%%%%

\addtocontents{toc}{\vspace{-1em}}
\section[\hspace{5em}Comparison to previous approaches]{Comparison to previous approaches} \label{app:comparison}

Here we check the consistency of the new semiclassical approach for the second moments with previous methods.

\subsection{The second moment $m_2$}

Previous approaches to the moments of the proper time-delays involved including an energy dependence during the intermediate semiclassical calculations which is differentiated out in a final step \cite{ks08,bk10,bk11}.
For example, by including an energy dependence in the scattering matrix and considering the correlation function
\begin{equation} \label{Cepsilon}
C(\epsilon) = \frac{1}{M} \Tr \left[S^{\dagger}\left(E-\frac{\epsilon\hbar\mu}{2}\right)S\left(E+\frac{\epsilon\hbar\mu}{2}\right)\right] \, ,
\end{equation}
the first moment can be obtained as follows
\begin{equation}
m_1 =  \frac{1}{\rmi \mu} \frac{\rmd}{\rmd \epsilon} C(\epsilon) \Big\vert_{\epsilon=0} \, ,
\end{equation}
in line with \eref{eq:Q}.
By treating correlated pairs of semiclassical trajectories, this was shown to equal the average time delay in \cite{ks08}.
For second moments one should in general consider a correlator of four (energy dependent) scattering matrices.
To avoid that here, we can actually use the unitarity of the scattering matrix to obtain \cite{bk12}
\begin{equation} \label{secmomCepsrel}
m_2 =  \frac{1}{(\rmi \mu)^2} \frac{\rmd^{2}}{\rmd \epsilon^{2}} C(\epsilon) \Big\vert_{\epsilon=0} \, ,
\end{equation}
and enormously reduce the complexity of the problem.  We state the semiclassical result for the correlator \cite{ks08}
\begin{equation} \label{Cepsilonsemi}
\fl C(\epsilon) = \left(1+\frac{2-\beta}{\beta M}\right) \sum_{K=0}^{\infty}\frac{1}{M^{K}}\sum_{v}^{L-V=K}(-1)^V\frac{\prod_{\sigma=1}^{V}(1-\rmi\epsilon l_{\sigma})}{(1-\rmi\epsilon)^{L+1}}N(v) \, ,
\end{equation}
in terms of a sum over periodic orbits (since these are cut once to obtain the first moment diagrams) and we include the diagonal term as a vector with 0 entries.
The prefactor accounts for the coherent backscattering diagrams:  With TRS, when the incoming and outgoing channel coincide, the time reversal of the partner trajectory can also be paired with the original trajectory.
Differentiating
\begin{equation}
\fl \frac{\rmd}{\rmd \epsilon}\frac{\prod_{\sigma=1}^{V}(1-\rmi\epsilon l_{\sigma})}{(1-\rmi\epsilon)^{L+1}} = \rmi \left[\frac{L+1}{(1-\rmi\epsilon)}-\sum_{\sigma}\frac{l_{\sigma}}{(1-\rmi\epsilon l_\sigma)}\right]\frac{\prod_{\sigma=1}^{V}(1-\rmi\epsilon l_{\sigma})}{(1-\rmi\epsilon)^{L+1}} \, ,
\end{equation}
\begin{equation}
\fl \frac{\rmd^2}{\rmd \epsilon^2}\frac{\prod_{\sigma=1}^{V}(1-\rmi\epsilon l_{\sigma})}{(1-\rmi\epsilon)^{L+1}}  \Big\vert_{\epsilon=0} = -\left[L+1-\sum_{\sigma}l_{\sigma}^{2}\right] - \left[L+1-\sum_{\sigma}l_{\sigma}\right]^{2} \, .
\end{equation}
Since $\sum_{\sigma}l_{\sigma}=\sum_{l}lv_l=L$ the second term is simply 1 and so
\begin{equation} \label{Cepsilonseminew}
\fl \frac{\rmd^{2}}{\rmd \epsilon^{2}} C(\epsilon)\Big\vert_{\epsilon=0} = -\left(1+\frac{2-\beta}{\beta M}\right) \sum_{K=0}^{\infty}\frac{1}{M^{K}}\sum_{v}^{L-V=K}(-1)^V\left[L+2-\sum_{l}l^2v_l\right]N(v) \, .
\end{equation}
These are sums we have already evaluated and hence
\begin{equation} \label{m2energy}
\mu^2 m_2 = \left(1+\frac{2-\beta}{\beta M}\right) \sum_{K=0}^{\infty}\frac{A_K+C_K-J_K}{M^{K}} \, .
\end{equation}

Without TRS, $\beta=2$, the sum $A_K+C_K-J_K=2$ for all even $K$ (including $K=0$) and hence
\begin{equation} \label{m2energyresultunit}
\mu^2 m_2 = \sum_{k=0}^{\infty}\frac{2}{M^{2k}} = \frac{2M^2}{M^2-1} \, ,
\end{equation}
in agreement with \eref{m2unitresult}.
With TRS, $\beta=1$, we have the sum
\begin{eqnarray} \label{m2energyresultortho}
\fl \mu^2 m_2 = 2 - \frac{1}{M}\sum_{K=0}^{\infty}\frac{I_K+I_{K+1}}{M^{K}} &=& 2 +\frac{2}{3M}\sum_{K=0}^{\infty}\frac{4\cdot2^{K}-(-1)^K}{M^{K}} \\
&=& \frac{2M^2}{(M+1)(M-2)} \, , \nonumber
\end{eqnarray}
in agreement with \eref{m2orthoresult}. Even for the unitary case we are forced to use the semiclassical sums from \ref{app:unitsums} and the calculation is only tractable in this way because we used \eref{secmomCepsrel}.
Nonetheless, this shows the agreement between the new semiclassical method presented here and previous approaches.

\subsection{The variance of the Wigner time delay}

For the variance of the Wigner time delay, to avoid treating energy dependent correlations between quadruplets of scattering trajectories, we turn to the expression in terms of periodic orbit correlations  like \eref{eq:timedelayvar} as discussed in \sref{intro}. In particular we may derive an expression for the two point correlator of the time delay at different energies as a second differential (\cf the appendix in the preprint version of \cite{mulleretal07}). This can only be done for the off-diagonal terms, while the diagonal term was given in \eref{eq:vartdleadingord}.
Here we merely state the semiclassical expression
\begin{equation} \label{vartdorbitssemi}
\fl \var(\tW) = \frac{4}{\beta M^2} - \frac{2}{\beta M^2} \frac{\rmd^{2}}{\rmd \epsilon^{2}}\sum_{K=1}^{\infty}\frac{1}{M^{K}}\sum_{v}^{L-V=K}(-1)^V\frac{\prod_{\sigma=1}^{V}(1-\rmi\epsilon l_{\sigma})}{L(1-\rmi\epsilon)^{L}}N(v) \Big\vert_{\epsilon=0}  \, .
\end{equation}
The main difference is that since periodic orbits are closed, we divide by the number of links $L$ to avoid overcounting the same orbits while in total there are $L$ links which is one fewer than when the periodic orbits are cut open to create the conductance diagrams used for \eref{Cepsilonsemi}.
With TRS we may also always compare an orbit with its correlated partner and its time reversal, giving the global factor of $\frac{2}{\beta}$.
Performing the differentials
\begin{equation}
\frac{\rmd^2}{\rmd \epsilon^2}\frac{\prod_{\sigma=1}^{V}(1-\rmi\epsilon l_{\sigma})}{(1-\rmi\epsilon)^{L}}  \Big\vert_{\epsilon=0} = -\left[L-\sum_{\sigma}l_{\sigma}^{2}\right] - \left[L-\sum_{\sigma}l_{\sigma}\right]^{2} \, ,
\end{equation}
the second term cancels completely and we are left with
\begin{equation} \label{vartdorbitssemisimp}
\fl \var(\tW) = \frac{4}{\beta M^2} + \frac{4}{\beta M^2} \sum_{K=1}^{\infty}\frac{1}{M^{K}}\sum_{v}^{L-V=K}(-1)^V \left[1-\frac{\sum_{l}l^2v_l}{L}\right]N(v)  \, ,
\end{equation}
again in terms of sums we know
\begin{equation} \label{vartorbitsdsemires}
\var(\tW) = \frac{4}{\beta M^2} + \frac{4}{\beta M^2} \sum_{K=1}^{\infty}\frac{C_K-H_K}{M^{K}}  \, .
\end{equation}
Without TRS, $C_K-H_K=1$ for all $K$ (including 0) and hence
\begin{equation} \label{vartdorbitsunitres}
\var(\tW) = \frac{2}{M^2}\sum_{k=0}^{\infty}\frac{1}{M^{2k}} = \frac{2}{M^2-1} \, ,
\end{equation}
the same as \eref{vartdunitresult}.  With TRS instead
\begin{equation} \label{vartdorbitsorthores}
\var(\tW) = \frac{4}{3M^2}\sum_{k=0}^{\infty}\frac{2\cdot 2^{K}+(-1)^K}{M^{2k}} = \frac{4}{(M+1)(M-2)} \, ,
\end{equation}
the same as \eref{vartdorthoresult}.

%%%%%%%%%%%%%%%%%%%%%%%%%%%%%%%%%%%%%%%%%%%%%%%%%

\addtocontents{toc}{\vspace{-1em}}
\section[\hspace{5em}Algorithmic approach to moment generating functions]{Algorithmic approach to moment generating functions} \label{app:algorithmic}

In \cite{bk13b} the algorithmic approach requires knowledge of the semiclassical contributions of the different types of edges and vertices.  The edges are matched together at the vertices to make all the permissible semiclassical diagrams at a given order in $M^{-1}$.  Already at subleading order in \sref{sec:subleadingmomgen}, we saw a single edge connected to itself to form a M\"obius strip.

The possible types of edges can be described by the types of leaves they would have at their ends, if they ended in leaves.  With TRS there are two types -- one involving an odd number of odd nodes
\begin{equation}
\fl E(oi,io)=E(io,oi)=\frac{1}{M}\frac{\mathcal{B}}{(1-\mathcal{A})^2-\mathcal{B}^2}
 = -\frac{1}{M}\frac{h^2}{h^2+2h-1} \, ,
\end{equation}
the other involving an even number.
\begin{equation}
\fl E(io,io)=E(oi,oi)=\frac{1}{M}\frac{1-\mathcal{A}}{(1-\mathcal{A})^2-\mathcal{B}^2}
 = \frac{1}{M}\frac{2h-1}{h^2+2h-1} \, ,
\end{equation}

For edges without TRS and traversed in the same direction on either side we also need to consider odd nodes with an excess of $f$
\begin{eqnarray}
\mathcal{B}_o &=& -\sum_{l=2}^{\infty} (l-1) f^2h^{l-2} + r f \sum_{l=2}^{\infty} l(l-1) h^{l-2} \\
&=& -\frac{f^2}{(1-h)^2}+\frac{2rf}{(1-h)^3} = \frac{f^2}{(1-h)^2} \, , \nonumber
\end{eqnarray}
or $\fh$ type subtrees
\begin{eqnarray}
\fl \mathcal{B}_i &=& -\sum_{l=2}^{\infty} (l-1) \fh^2h^{l-2} + r \fh^{3} \sum_{l=2}^{\infty} (l-1)(l-2) h^{l-3} + \sum_{l=2}^{\infty}(l-1)r^lf^{l-2}\nonumber \\
\fl &=& -\frac{\fh^2}{(1-h)^2}+\frac{2r\fh^3}{(1-h)^3} +\frac{r^2}{(1-rf)^2}  = \frac{\fh^2(2h-1)}{(1-h)^2} +\fh^{2} = \frac{\fh^2h^2}{(1-h)^2} \, ,
\end{eqnarray}
which can also touch the incoming lead.

The edge contributions are then \cite{bk13b}
\begin{equation}
E(o,i)=E(i,o)=\frac{1}{M}\frac{(1-\mathcal{A})}{(1-\mathcal{A})^2-\mathcal{B}_o\mathcal{B}_i}
 = \frac{1}{M}\frac{2h-1}{h^2+2h-1} \, ,
\end{equation}
\begin{equation}
E(o,o)=\frac{1}{M}\frac{\mathcal{B}_i}{(1-\mathcal{A})^2-\mathcal{B}_o\mathcal{B}_i}
 = -\frac{1}{M}\frac{\fh^2h^2}{h^2+2h-1} \, ,
\end{equation}
\begin{equation}
E(i,i)=\frac{1}{M}\frac{\mathcal{B}_o}{(1-\mathcal{A})^2-\mathcal{B}_o\mathcal{B}_i}
 = -\frac{1}{M}\frac{f^2}{h^2+2h-1} \, .
\end{equation}

For the vertices of degree $k$, we label the edge stumps by the components of a vector $\bb$.  If adjoining components are identical, we need an even number of subtrees in that sector, otherwise an odd number with an appropriate excess of one type of subtree $f$ or $\fh$.  The normal contribution is
\begin{equation}
\tilde{V}_k(\bb) = -M \frac{f^q\fh^p}{(1-h)^k} \, ,
\end{equation}
where $q$ is the number of times $i$ follows $i$ in the sequence $\bb$ (taken cyclically) and $p$ is the number of times $o$ follows $o$.  Next, any of the $f$ type trees can connect directly to an outgoing leaf, which we can move into the encounter at the vertex.  This give a further contribution of
\begin{equation}
-r\frac{\partial}{\partial f}\tilde{V}_k(\bb) = -\tilde{V}_k(\bb)\left[q(1-h)+kh\right] \, .
\end{equation}
Finally, if all the sectors are odd with an excess of $\fh$ types subtrees so that $p=k$, the encounter can move into the lead giving an extra contribution of
\begin{equation}
\left(\frac{r}{1-rf}\right)^k = \fh^k \, .
\end{equation}
All combined we have
\begin{equation}
V_k(\bb) = M \frac{f^q\fh^p}{(1-h)^k}\left[-1+q(1-h)+kh\right] +M\delta_{p,k}\fh^k \, .
\end{equation}

These contributions can be plugged into the algorithm of \cite{bk13b} to give the moment generating functions detailed in \sref{sec:algoresults}.

%%%%%%%%%%%%%%%%%%%%%%%%%%%%%%%%%%%%%%%%%%%%%%%%%%%%%%%%

\section*{References}
%\bibliography{esadt}

\begin{thebibliography}{100}

\bibitem{eisenbud48}
L.~Eisenbud
\newblock 1948
\newblock {\em The formal properties of nuclear collisions}
\newblock PhD thesis, Princeton University

\bibitem{wigner55}
E.~P. Wigner
\newblock 1955
\newblock {\em Phys. Rev.}, {\textbf{98}} 145--147

\bibitem{smith60}
F.~T. Smith
\newblock 1960
\newblock {\em Phys. Rev.}, {\textbf{118}} 349--356

\bibitem{lyuboshitz77}
V.~L. Lyuboshitz
\newblock 1977
\newblock {\em Phys. Lett. B}, {\textbf{72}} 41--44

\bibitem{lewe91}
C.~H. Lewenkopf and H.~A. Weidenm\"{u}ller
\newblock 1991
\newblock {\em Ann. Phys.}, {\textbf{212}} 53

\bibitem{lehmannetal95}
N.~Lehmann, D.~V. Savin, V.~V. Sokolov and H.-J. Sommers
\newblock 1995
\newblock {\em Physica D}, {\textbf{86}} 572--585

\bibitem{fs97}
Y.~V. Fyodorov and H.~{\relax -J}. Sommers
\newblock 1997
\newblock {\em J. Math. Phys.}, {\textbf{38}} 1918--1981

\bibitem{muga00}
J.~G. Muga and C.~R. Leavens
\newblock 2000
\newblock {\em Phys. Rep.}, {\textbf{338}} 353

\bibitem{deCar02}
C.~A.~A. de~Carvalho and H.~M. Nussenzveig
\newblock 2002
\newblock {\em Phys. Rep.}, {\textbf{364}} 83

\bibitem{friedel52}
J.~Friedel
\newblock 1952
\newblock {\em Phil. Mag.}, {\textbf{43}} 153--189

\bibitem{krei53}
M.~G. Krein
\newblock 1953
\newblock {\em Mat. Sbornik}, {\textbf{33}} 597--626

\bibitem{Stoeckmann}
H.-J. St{\"{o}}ckmann
\newblock 1999
\newblock {\em Quantum Chaos: An Introduction}
\newblock Cambridge University Press, Cambridge, UK

\bibitem{alha00}
Y.~Alhassid
\newblock 2000
\newblock {\em Rev. Mod. Phys.}, {\textbf{72}} 895

\bibitem{mitc10}
G.~E. Mitchell, A.~Richter and H.~A. Weidenm\"uller
\newblock 2010
\newblock {\em Rev. Mod. Phys.}, {\textbf{82}} 2845--2901

\bibitem{Mehta2}
M.~L. Mehta
\newblock 1991
\newblock {\em Random Matrices}
\newblock Academic Press, New York, 2nd edition

\bibitem{gutzwiller90}
M.~Gutzwiller
\newblock 1990
\newblock {\em Chaos in Classical and Quantum Mechanics}
\newblock Springer, New York

\bibitem{sss01}
{H.-J}. Sommers, D.~V. Savin and V.~V. Sokolov
\newblock 2001
\newblock {\em Phys. Rev. Lett.}, {\textbf{87}} 094101

\bibitem{soko89}
V.~V. Sokolov and V.~G. Zelevinsky
\newblock 1989
\newblock {\em Nucl. Phys. A}, {\textbf{504}} 562

\bibitem{verb85}
J.~J.~M. Verbaarschot, H.~A. Weidenm{\"{u}}ller and M.~R. Zirnbauer
\newblock 1985
\newblock {\em Phys. Rep.}, {\textbf{129}} 367

\bibitem{guhr98}
T.~Guhr, A.~{M\"{u}ller-Groeling} and H.~A. Weidenm\"{u}ller
\newblock 1998
\newblock {\em Phys. Rep.}, {\textbf{299}} 189

\bibitem{fyod97a}
Y.~V. Fyodorov, D.~V. Savin and H.-J. Sommers
\newblock 1997
\newblock {\em Phys. Rev. E}, {\textbf{55}} R4857

\bibitem{savi01}
D.~V. Savin, Y.~V. Fyodorov and H.-J. Sommers
\newblock 2001
\newblock {\em Phys. Rev. E}, {\textbf{63}} 035202

\bibitem{ss03}
D.~V. Savin and {H.-J}. Sommers
\newblock 2003
\newblock {\em Phys. Rev. E}, {\textbf{68}} 036211

\bibitem{fyod05}
Y.~V. Fyodorov, D.~V. Savin and H.-J. Sommers
\newblock 2005
\newblock {\em J. Phys. A}, {\textbf{38}} 10731

\bibitem{ossi05}
A.~Ossipov and Y.~V. Fyodorov
\newblock 2005
\newblock {\em Phys. Rev. B}, {\textbf{71}} 125133

\bibitem{fyod11ox}
Y.~V. Fyodorov and D.~V. Savin \newblock 2011 
\newblock Resonance scattering of waves in chaotic systems
\newblock {\em The Oxford Handbook of Random Matrix Theory} editors G.~Akemann, J.~Baik and P.~Di~Francesco, (Oxford University Press) pp 703--722 [arXiv:1003.0702]

\bibitem{beenakker97}
C.~W.~J. Beenakker
\newblock 1997
\newblock {\em Rev. Mod. Phys.}, {\textbf{69}} 731--808

\bibitem{bfb97}
P.~W. Brouwer, K.~M. Frahm and C.~W.~J. Beenakker
\newblock 1997
\newblock {\em Phys. Rev. Lett.}, {\textbf{78}} 4737--4740

\bibitem{gopa96}
V.~A. Gopar, P.~A. Mello and M.~B\"uttiker
\newblock 1996
\newblock {\em Phys. Rev. Lett.}, {\textbf{77}} 3005

\bibitem{bfb99}
P.~W. Brouwer, K.~M. Frahm and C.~W.~J. Beenakker
\newblock 1999
\newblock {\em Waves in Random Media}, {\textbf{9}} 91--104

\bibitem{mezz11}
F.~Mezzadri and N.~Simm
\newblock 2011
\newblock {\em J. Math. Phys.}, {\textbf{52}} 103511

\bibitem{mezz12}
F.~Mezzadri and N.~Simm
\newblock 2012
\newblock {\em J. Math. Phys.}, {\textbf{53}} 053504

\bibitem{mart14}
A.~M. Mart\'inez-Arg\"uello, M.~Mart\'inez-Mares and J.~C. Garc\'ia
\newblock 2014
\newblock {\em J. Math. Phys}, {\textbf{55}} 081901

\bibitem{nova14}
M.~Novaes
\newblock 2014
\newblock Preprint, {a}rXiv:1408.1669

\bibitem{texi13}
C.~Texier and S.~N. Majumdar
\newblock 2013
\newblock {\em Phys. Rev. Lett.}, {\textbf{110}} 250602

\bibitem{mezz13}
F.~Mezzadri and N.~Simm
\newblock 2013
\newblock {\em Comm. Math. Phys.}, {\textbf{324}} 465--513

\bibitem{gopa98}
V.~A. Gopar and P.~A. Mello
\newblock 1998
\newblock {\em Europhys. Lett.}, {\textbf{42}} 131

\bibitem{gutzwiller71}
M.~C. Gutzwiller
\newblock 1971
\newblock {\em J. Math. Phys.}, {\textbf{12}} 343--358

\bibitem{bb74}
R.~Balian and C.~Bloch
\newblock 1974
\newblock {\em Ann. Phys.}, {\textbf{85}} 514--545

\bibitem{eckh93}
B.~Eckhardt
\newblock 1993
\newblock {\em Chaos}, {\textbf{3}} 613--617

\bibitem{val98}
R.~O. Vallejos, A.~M. Ozorio~de Almeida and C.~H. Lewenkopf
\newblock 1998
\newblock {\em J. Phys. A}, {\textbf{31}} 4885--4897

\bibitem{berry85}
M.~V. Berry
\newblock 1985
\newblock {\em Proc. Roy. Soc. A}, {\textbf{400}} 229--251

\bibitem{ce91}
P.~Cvitanovi\'c and B.~Eckhardt
\newblock 1991
\newblock {\em J. Phys. A}, {\textbf{24}} L237--L241

\bibitem{ha84}
J.~H. Hannay and A.~M. Ozorio~de Almeida
\newblock 1984
\newblock {\em J. Phys. A}, {\textbf{17}} 3429--3440

\bibitem{ss97}
D.~V. Savin and V.~V. Sokolov
\newblock 1997
\newblock {\em Phys. Rev. E}, {\textbf{56}} R4911--R4913

\bibitem{lv04}
C.~H. Lewenkopf and R.~O. Vallejos
\newblock 2004
\newblock {\em J. Phys. A}, {\textbf{37}} 131--136

\bibitem{lehm95a}
N.~Lehmann, D.~Saher, V.~V. Sokolov and H.-J. Sommers
\newblock 1995
\newblock {\em Nucl. Phys. A}, {\textbf{582}} 223

\bibitem{sr01}
M.~Sieber and K.~Richter
\newblock 2001
\newblock {\em Phys. Scr.}, {\textbf{T90}} 128--133

\bibitem{mulleretal04}
S.~M\"uller, S.~Heusler, P.~Braun, F.~Haake and A.~Altland
\newblock 2004
\newblock {\em Phys. Rev. Lett.}, {\textbf{93}} 014103

\bibitem{mulleretal05}
S.~M\"uller, S.~Heusler, P.~Braun, F.~Haake and A.~Altland
\newblock 2005
\newblock {\em Phys. Rev. E}, {\textbf{72}} 046207

\bibitem{ks07b}
J.~Kuipers and M.~Sieber
\newblock 2007
\newblock {\em Nonlinearity}, {\textbf{20}} 909--926

\bibitem{heus07}
S.~Heusler, S.~M\"{u}ller, A.~Atland, P.~Braun and F.~Haake
\newblock 2007
\newblock {\em Phys. Rev. Lett.}, {\textbf{98}} 044103

\bibitem{muel09}
S.~M{\"{u}}ller, S.~Heusler, A.~Altland, P.~Braun and F.~Haake
\newblock 2009
\newblock {\em New J. Phys.}, {\textbf{9}} 103025

\bibitem{miller75}
W.~H. Miller
\newblock 1975
\newblock {\em Adv. Chem. Phys.}, {\textbf{30}} 77--136

\bibitem{jbs90}
R.~A. Jalabert, H.~U. Baranger and A.~D. Stone
\newblock 1990
\newblock {\em Phys. Rev. Lett.}, {\textbf{65}} 2442--2445

\bibitem{bjs93b}
H.~U. Baranger, R.~A. Jalabert and A.~D. Stone
\newblock 1993
\newblock {\em Chaos}, {\textbf{3}} 665--682

\bibitem{richter00}
K.~Richter
\newblock 2000
\newblock {\em Semiclassical theory of mesoscopic quantum systems}
\newblock Springer, Berlin

\bibitem{rs02}
K.~Richter and M.~Sieber
\newblock 2002
\newblock {\em Phys. Rev. Lett.}, {\textbf{89}} 206801

\bibitem{heusleretal06}
S.~Heusler, S.~M\"uller, P.~Braun and F.~Haake
\newblock 2006
\newblock {\em Phys. Rev. Lett.}, {\textbf{96}} 066804

\bibitem{mulleretal07}
S.~M\"uller, S.~Heusler, P.~Braun and F.~Haake
\newblock 2007
\newblock {\em New J. Phys.}, {\textbf{9}} 12
\newblock Additional appendices in preprint version arXiv:cond-mat/0610560.

\bibitem{ks08}
J.~Kuipers and M.~Sieber
\newblock 2008
\newblock {\em Phys. Rev. E}, {\textbf{77}} 046219

\bibitem{bk10}
G.~Berkolaiko and J.~Kuipers
\newblock 2010
\newblock {\em J. Phys. A}, {\textbf{43}} 035101

\bibitem{bk11}
G.~Berkolaiko and J.~Kuipers
\newblock 2011
\newblock {\em New J. Phys}, {\textbf{13}} 063020

\bibitem{sz97}
V.~V. Sokolov and V.~Zelevinsky
\newblock 1997
\newblock {\em Phys. Rev. C}, {\textbf{56}} 311--323

\bibitem{bb00}
{\mbox{Ya}}.~M. Blanter and M.~B{\"{u}}ttiker
\newblock 2000
\newblock {\em Phys. Rep.}, {\textbf{336}} 1--166

\bibitem{ss06}
D.~V. Savin and {H.-J}. Sommers
\newblock 2006
\newblock {\em Phys. Rev. B}, {\textbf{73}} 081307

\bibitem{somm07a}
H.-J. Sommers, W.~Wieczorek and D.~V. Savin
\newblock 2007
\newblock {\em Acta Phys. Pol. A}, {\textbf{112}} 691

\bibitem{novaes07}
M.~Novaes
\newblock 2007
\newblock {\em Phys. Rev. B}, {\textbf{75}} 073304

\bibitem{ssw08}
D.~V. Savin, {H.-J}. Sommers and W.~Wieczorek
\newblock 2008
\newblock {\em Phys. Rev. B}, {\textbf{77}} 125332

\bibitem{novaes08}
M.~Novaes
\newblock 2008
\newblock {\em Phys. Rev. B}, {\textbf{78}} 035337

\bibitem{kss09}
B.~A. Khoruzhenko, D.~V. Savin and {H.-J}. Sommers
\newblock 2009
\newblock {\em Phys. Rev. B}, {\textbf{80}} 125301

\bibitem{vv08}
P.~Vivo and E.~Vivo
\newblock 2008
\newblock {\em J. Phys. A}, {\textbf{41}} 122004

\bibitem{ok08}
V.~A. Osipov and E.~Kanzieper
\newblock 2008
\newblock {\em Phys. Rev. Lett.}, {\textbf{101}} 176804

\bibitem{ok09}
V.~A. Osipov and E.~Kanzieper
\newblock 2009
\newblock {\em J. Phys. A}, {\textbf{42}} 475101

\bibitem{liva11}
G.~Livan and P.~Vivo
\newblock 2011
\newblock {\em Acta Phys. Pol. B}, {\textbf{42}} 1081--1104

\bibitem{souz14}
C.~A. {Souza-Filho}, A.~F. Macedo-Junior and A.~M.~S. Mac\^{e}do
\newblock 2014
\newblock {\em J. Phys. A}, {\textbf{47}} 105102

\bibitem{bhn08}
G.~Berkolaiko, J.~M. Harrison and M.~Novaes
\newblock 2008
\newblock {\em J. Phys. A}, {\textbf{41}} 365102

\bibitem{pich01}
K.~Pichugin, H.~Schanz and P.~Seba
\newblock 2001
\newblock {\em Phys. Rev. E}, {\textbf{64}} 056227

\bibitem{stoec02i}
H.-J. St{\"{o}}ckmann, E.~Persson, Y.-H. Kim, M.~Barth, U.~Kuhl and I.~Rotter
\newblock 2002
\newblock {\em Phys. Rev. E}, {\textbf{65}} 066211

\bibitem{savi03}
D.~V. Savin, V.~V. Sokolov and H.-J. Sommers
\newblock 2003
\newblock {\em Phys. Rev. E}, {\textbf{67}} 026215

\bibitem{fl81}
D.~S. Fisher and P.~A. Lee
\newblock 1981
\newblock {\em Phys. Rev. B}, {\textbf{23}} 6851--6854

\bibitem{kuipersetal09}
J.~Kuipers, D.~Waltner, M.~Guti\'errez and K.~Richter
\newblock 2009
\newblock {\em Nonlinearity}, {\textbf{22}} 1945--1964

\bibitem{waltneretal08}
D.~Waltner, M.~Guti\'errez, A.~Goussev and K.~Richter
\newblock 2008
\newblock {\em Phys. Rev. Lett.}, {\textbf{101}} 174101

\bibitem{gutierrezetal09}
M.~Guti\'errez, D.~Waltner, J.~Kuipers and K.~Richter
\newblock 2009
\newblock {\em Phys. Rev. E}, {\textbf{79}} 046212

\bibitem{gutkinetal10}
B.~Gutkin, D.~Waltner, M.~Guti\'errez, J.~Kuipers and K.~Richter
\newblock 2010
\newblock {\em Phys. Rev. E}, {\textbf{81}} 036222

\bibitem{sieber99}
M.~Sieber
\newblock 1999
\newblock {\em J. Phys. A}, {\textbf{32}} 7679--7689

\bibitem{kuipersetal11}
J.~Kuipers, T.~Engl, G.~Berkolaiko, C.~Petitjean, D.~Waltner and K.~Richter
\newblock 2011
\newblock {\em Phys. Rev. B}, {\textbf{83}} 195316

\bibitem{bk12}
G.~Berkolaiko and J.~Kuipers
\newblock 2012
\newblock {\em Phys. Rev. E}, {\textbf{85}} 045201

\bibitem{bk13a}
G.~Berkolaiko and J.~Kuipers
\newblock 2013
\newblock {\em J. Math. Phys.}, {\textbf{54}} 112103

\bibitem{bk13b}
G.~Berkolaiko and J.~Kuipers
\newblock 2013
\newblock {\em J. Math. Phys.}, {\textbf{54}} 123505

\bibitem{braunetal06}
P.~Braun, S.~Heusler, S.~M\"uller and F.~Haake
\newblock 2006
\newblock {\em J. Phys. A}, {\textbf{39}} L159--L165

\bibitem{bb96}
P.~W. Brouwer and C.~W.~J. Beenakker
\newblock 1996
\newblock {\em J. Math. Phys.}, {\textbf{37}} 4904--4934

\bibitem{matsumoto12}
S.~Matsumoto
\newblock 2012
\newblock {\em J. Theor. Prob.}, {\textbf{25}} 798--822

\bibitem{nova13c}
M.~Novaes
\newblock 2013
\newblock {\em J. Phys. A}, {\textbf{46}} 502002

\bibitem{marc14}
M.~Marciani, P.~W. Brouwer and C.~W.~J. Beenakker
\newblock 2014
\newblock {\em Phys. Rev. B}, {\textbf{90}} 045403

\bibitem{alei96}
I.~L. Aleiner and A.~I. Larkin
\newblock 1996
\newblock {\em Phys. Rev. B}, {\textbf{54}} 14423--14444

\bibitem{ada03}
I.~Adagideli
\newblock 2003
\newblock {\em Phys. Rev. B}, {\textbf{68}} 233308

\bibitem{jw06}
{\relax Ph}.~Jacquod and R.~S. Whitney
\newblock 2006
\newblock {\em Phys. Rev. B}, {\textbf{73}} 195115

\bibitem{br06a}
P.~W. Brouwer and S.~Rahav
\newblock 2006
\newblock {\em Phys. Rev. B}, {\textbf{74}} 075322

\bibitem{wj06}
R.~S. Whitney and {\relax Ph}.~Jacquod
\newblock 2006
\newblock {\em Phys. Rev. Lett.}, {\textbf{96}} 206804

\bibitem{wkr11}
D.~Waltner, J.~Kuipers and K.~Richter
\newblock 2011
\newblock {\em Phys. Rev. B}, {\textbf{83}} 195315

\bibitem{wkjr12}
D.~Waltner, J.~Kuipers, {\relax Ph}.~Jacquod and K.~Richter
\newblock 2012
\newblock {\em Phys. Rev. B}, {\textbf{85}} 024302

\bibitem{kr13}
J.~Kuipers and K.~Richter
\newblock 2013
\newblock {\em J. Phys. A}, {\textbf{46}} 055101

\bibitem{mell90}
P.~A. Mello
\newblock 1990
\newblock {\em J. Phys. A}, {\textbf{23}} 4061

\bibitem{zfr05}
O.~Zaitsev, D.~Frustaglia and K.~Richter
\newblock 2005
\newblock {\em Phys. Rev. Lett.}, {\textbf{94}} 026809

\bibitem{zfr05b}
O.~Zaitsev, D.~Frustaglia and K.~Richter
\newblock 2005
\newblock {\em Phys. Rev. B}, {\textbf{72}} 155325

\bibitem{wb07}
D.~Waltner and J.~Bolte
\newblock 2007
\newblock {\em Phys. Rev. B}, {\textbf{76}} 075330

\bibitem{mell80}
P.~A. Mello and T.~Seligman
\newblock 1980
\newblock {\em Nucl. Phys. A}, {\textbf{344}} 489

\end{thebibliography}

\end{document}